\documentclass[useAMS,usenatbib]{mn2e}
\usepackage{epsfig}
\usepackage{float}

\usepackage{natbib}

\usepackage{color}

\usepackage{aas_macros}

\def\sigc{\mbox{$\sigma_0$}}
\def\sigs{\mbox{$\sigma_\star$}}
\def\sige{\mbox{$\sigma_{\rm e}$}}
\def\vmax{\mbox{$V_{\rm max}$}}

\def\Re{\mbox{$R_{\rm e}$}}

\def\Msun{\mbox{$M_\odot$}}
\def\Mtot{\mbox{$M_{\rm tot}$}}

\def\ML{\mbox{$M/L$}}
\def\dimf{\mbox{$\delta_{\rm IMF}$}}
\def\Yst{\mbox{$\Upsilon_\star$}}

\def\Mdyn{\mbox{$M_{\rm dyn}$}}

\def\mst{\mbox{$M_{\star}$}}

\def\Mvir{\mbox{$M_{\rm vir}$}}

\def\fdm{\mbox{$f_{\rm DM}$}}

\def\lsim{\mathrel{\rlap{\lower3.5pt\hbox{\hskip0.5pt$\sim$}}
    \raise0.5pt\hbox{$<$}}}                
\def\gsim{~\rlap{$>$}{\lower 1.0ex\hbox{$\sim$}}}

\def\sigAp{\mbox{$\sigma_{\rm Ap}$}}
\def\TtoSM{\mbox{$M_{\rm tot}/\mst$}}

\def\atlas3d{ATLAS$^{\rm 3D}$}

\title[\fdm\ vs z]{Evolution of central dark matter of early-type galaxies up to $z \sim 0.8$}

\author[Tortora C. et al.]{\noindent
C.~Tortora$^{1}$\thanks{E-mail: ctortora@na.astro.it},
N.R.~Napolitano$^{1}$, R. P.~Saglia$^{2,3}$,
A.J.~Romanowsky$^{4,5}$, G.~Covone$^{6,7}$, \and
M.~Capaccioli$^{6,7}$
\\~\\
$^1$ INAF -- Osservatorio Astronomico di
Capodimonte, Salita Moiariello, 16, 80131 - Napoli, Italy\\
$^2$ Max-Planck-Institut f$\ddot{u}$r extraterrestrische
Physik, Giessenbachstrasse, D-85748 Garching, Germany\\
$^3$ Universit$\ddot{a}$ts-Sternwarte M$\ddot{u}$nchen,
Scheinerstrasse 1, D-81679 M$\ddot{u}$nchen, Germany\\
$^4$ Department of Physics and Astronomy, San Jos\'e State
University, San Jose, CA 95192, USA\\
$^5$ University of California Observatories, 1156 High
Street, Santa Cruz, CA 95064, USA \\
$^6$ Dipartimento di Scienze Fisiche, Universit\`{a} di Napoli
Federico II, Compl. Univ. Monte S. Angelo, 80126 - Napoli, Italy\\
$^7$ INFN Sez. di Napoli, Compl. Univ. Monte S. Angelo, Via
Cinthia, I-80126 Napoli, Italy\\}

\begin{document}
\date{Accepted  Received }
\pagerange{\pageref{firstpage}--\pageref{lastpage}} \pubyear{xxxx}
\maketitle

\label{firstpage}
\begin{abstract}
We investigate the evolution of dark and luminous matter in the
central regions of early-type galaxies (ETGs) up to $z\sim 0.8$.
We use a spectroscopically selected sample of 154 cluster and
field galaxies from the EDisCS survey, covering a wide range in
redshifts ($z \sim$\,0.4--0.8), stellar masses ($\log \mst/\Msun
\sim$\, 10.5--11.5 dex) and velocity dispersions ($\sigs \sim$\,
100--300 \, km/s). We obtain central dark matter (DM) fractions by
determining the dynamical masses from Jeans modelling of galaxy
aperture velocity dispersions and the \mst\ from galaxy colours,
and compare the results with local samples. We discuss how the
correlations of central DM with galaxy size (i.e. the effective
radius, \Re), \mst\ and \sigs\ evolve as a function of redshift,
finding clear indications that local galaxies are, on average,
more DM dominated than their counterparts at larger redshift. This
DM fraction evolution with $z$ can be only partially interpreted
as a consequence of the size--redshift evolution. We discuss our
results within galaxy formation scenarios, and conclude that the
growth in size and DM content which we measure within the last 7
Gyr is incompatible with passive evolution, while it is well
reproduced in the multiple minor merger scenario. We also discuss
the impact of the IMF on our DM inferences and argue that this can
be non-universal with the lookback time. In particular, we find
the Salpeter IMF can be better accommodated by low redshift
systems, while producing stellar masses at high--$z$ which are
unphysically larger than the estimated dynamical masses
(particularly for lower-\sigs\ systems).
\end{abstract}

\begin{keywords}
galaxies: evolution  -- galaxies: general -- galaxies: elliptical
and lenticular, cD.
\end{keywords}

\section{Introduction}\label{sec:intro}

Dark matter (DM) is a ubiquitous component of the universe. It
dominates the mass density of virialized objects and has left its
imprint at cosmological scales during the cosmic history (e.g.,
\citealt{Komatsu+11_WMAP7}). In the last decade, the large wealth
of data from large sky surveys such as, e.g., the Sloan Digital
Sky Survey (SDSS, \citealt{SDSS_DR1,SDSS_DR6,SDSS_DR7}) have
established that the DM budget is up to $\sim 85\%$ of the total
mass density of the universe. Numerical simulations of (DM only)
structure formation within the consensus cosmology framework, i.e.
the $\Lambda$CDM model, have provided accurate predictions on the
DM density distribution in galaxies and clusters (\citealt{NFW96},
hereafter NFW; \citealt{Bullock+01}; \citealt{Maccio+08}). More
realistic models, which have tried to evaluate the effect of
baryons on the DM distribution, have been characterized either
through some approximate recipes such as the adiabatic contraction
(e.g., \citealt{Gnedin+04}), which seems compatible with
observations (e.g. \citealt{Gnedin+07}; \citealt{NRT10}) or to the
simultaneous evolutions of the two components (\citealt{Wu+14}),
making predictions on the expected DM fractions in the central
galaxy regions \citep{Hilz+13}.

Within this context, the study of the mass density distribution
and the dark-to-baryonic mass ratio of early-type galaxies (ETGs,
ellipticals and lenticulars) is a crucial problem that has
to be addressed in a systematic way, as these systems contain most of the cosmic stellar
mass of the universe and represent the final stage of galaxy evolution and
contain the fossil record of the stellar and DM assembly through time.

Most of the current understanding of these systems is based on local
galaxy samples. These have shown that the overall DM
content depends crucially on the galaxy mass scale, which is
possibly the effect of the overall star formation efficiency
(\citealt{Benson+00}; \citealt{MH02}; \citealt{Napolitano+05};
\citealt{Mandelbaum+06}; \citealt{vdB+07}; \citealt{CW09};
\citealt{Moster+10}).
DM has been found to be relevant in the central galaxy regions
(\citealt{Gerhard+01}; \citealt{Padmanabhan+04};
\citealt{Cappellari+06}; \citealt{ThomasJ+07};
\citealt{Cardone+09}; \citealt{ThomasJ+09}; \citealt{HB09_FP};
\citealt[T+09, hereafter]{Tortora+09}; \citealt{Auger+10_SLACSX};
\citealt{CT10}; \citealt{ThomasJ+11}; \citealt{Cardone+11SIM};
\citealt{SPIDER-VI}) where ETGs show a substantial consistency
with the concordance $\Lambda$CDM scenario (\citealt{NRT10};
\citealt{SPIDER-VI}) and also some new scaling relations, like the
one between DM fraction and formation epoch (\citealt{NRT10};
\citealt{Tortora+10lensing}), which still need to be confirmed and
fully understood.

Generally speaking, it has been shown that the central DM fraction
(typically within one effective radius, \Re\ hereafter) is larger
in more massive and large--sized galaxies (e.g. \citealt{HB09_FP};
T+09; \citealt{RS09}; \citealt{Auger+10_SLACSX}; \citealt{NRT10};
\citealt{ThomasJ+11}; \citealt{SPIDER-VI}), even though there are
also opposite claims from other studies (e.g.,
\citealt{Grillo+09}; \citealt{Grillo10};
\citealt{Grillo_Cobat10}). Although these trends are qualitatively
unchanged independently of the adopted galaxy model or initial
mass function, IMF (e.g., \citealt{Cardone+09}; \citealt{CT10};
\citealt{Cardone+11SIM}), they can still be strongly affected by
the assumptions on the stellar \ML, i.e. a non-homologous constant
profile, as verified for the trend of the DM fraction as a
function of mass (e.g. \citealt{TBB04}; T+09;
\citealt{SPIDER-VI}). The inventory of the evidences accumulated
so far is complicated if one takes into account the effect of a
non universal IMF (\citealt{Treu+10}; \citealt{ThomasJ+11};
\citealt{Conroy_vanDokkum12b}; \citealt{Cappellari+12,
Cappellari+13_ATLAS3D_XX}; \citealt{Spiniello+12};
\citealt{Wegner+12}; \citealt{Dutton+13}; \citealt{Ferreras+13};
\citealt{Goudfrooij_Kruijssen13};
\citealt{LaBarbera+13_SPIDERVIII_IMF}; \citealt{TRN13_SPIDER_IMF};
\citealt{Weidner+13_giant_ell}; \citealt{Tortora+14_MOND};
\citealt{Goudfrooij_Kruijssen14}), which remains the largest
source of uncertainty, since a systematic variation of the IMF
with mass, from a bottom-lighter IMF for low mass systems to a
bottom-heavier IMF in massive galaxies would wipe out the
``apparent'' DM fraction trend with mass (e.g.,
\citealt{ThomasJ+11}; \citealt{TRN13_SPIDER_IMF}).

A robust assessment of all these correlations is crucial to have a
clearer insight into the assembly of the two main galaxy
components (stars and DM), and thus into the formation mechanisms
across time. Unfortunately, with the currently available
individual datasets, which generally cover small windows in
redshift space, it is not possible to extend the investigation of
the scaling relations found at lower$-z$ to earlier epochs (but
see, for instance, \citealt{Auger+09_SLACSIX, Auger+10_SLACSX},
\citealt{Tortora+10lensing}; \citealt{Sonnenfeld+13_SL2S_IV} for
gravitational lenses). Only by accumulating data from different
samples, a more systematic study of the DM fraction evolution with
redshift is becoming possible.

This has been done, e.g., using  weak and/or strong lensing
analyses (e.g. \citealt{Heymans+06}; \citealt{Lagattuta+10}) to
find that at higher redshift results point to an evolution of the
total virial-to-stellar ratio that is larger at higher redshift.
For the central DM content, results are still inconclusive, as
both lower DM fractions (\citealt{Faure+11}, within the Einstein
radius) or larger ones (\citealt{Ruff+11}, within $\Re/2$) have
been found. However, the latter studies are based on limited
galaxy samples with different model choices (e.g. on adopted
scales at which the DM fractions are derived), thus their
conclusions are prone to uncertainties which are difficult to keep
under control.

Instead, homogeneous approaches on well studied high$-z$ samples
are still missing. Only recently, using massive galaxies from
SDSS-III/BOSS combined with a sample from SDSS-II,
\cite{Beifiori+14} have addressed the question, providing
evidences that high-z ETGs are less DM dominated than their local
counterparts. However, further independent analysis are needed, to
constrain not only the overall evolution of central DM, but also
how it correlates with mass or galaxy size, and how these
correlations change as a function of redshift.

The EDisCS sample (\citealt{White+05}; \citealt{Saglia+10})
includes ETGs in a wide range of redshifts ($\sim 0.4-0.8$) for
which accurate photometry, structural parameters and central
velocity dispersions are available. This provides us a rare
opportunity to investigate the evolution of the central DM
content, comparing these results with inferences in local galaxy
samples.

The paper is organized as follows. Data samples and the analysis
performed are introduced in Sect.~\ref{sec:data}. The evolution of
the relations between size and velocity dispersion with stellar
mass are discussed in Sect~\ref{sec:correlations}, while
Sect.~\ref{sec:evolution} is devoted to the analysis of central DM
content, its evolution with redshift, systematics and the
interpretation within the formation scenarios. Conclusions and
future prospects are discussed in Sect.~\ref{sec:conclusions}. We
adopt a cosmological model with
$(\Omega_{m},\Omega_{\Lambda},h)=(0.3,0.7,0.75)$, where $h =
H_{0}/100 \, \textrm{km} \, \textrm{s}^{-1} \, \textrm{Mpc}^{-1}$
(\citealt{Komatsu+11_WMAP7}).

\section{Samples and data analysis}\label{sec:data}

The aim of this paper is to present a uniform dynamical analysis
for galaxies distributed on a wide redshift window. As
higher-redshift galaxies we will use the sample from EDisCS survey
(\citealt{Saglia+10}), with covers a redshift window from
$z\sim0.4$ to $z\sim0.8$. As $z\sim0$ comparison samples, we use
the data from SPIDER project (\citealt{SPIDER-I}), \atlas3d\
(\citealt{Cappellari+11_ATLAS3D_I}) and from \cite{Tortora+09}. In
the following we will provide further details about data samples
adopted.

\subsection{EDisCS sample: data}

The EDisCS survey (\citealt{White+05}; \citealt{Saglia+10})
provides photometric and spectroscopic data for ETGs in field and
rich clusters with $0.4 \lsim z \lsim 0.8$ and stellar mass
completeness limit at $\log \mst/\Msun = 10.4$. We limit the
analysis to objects with spectroscopic measurements, as these
provide us accurate redshifts and internal kinematics. The final
sample which we will use for this analysis has been further
selected to have weak $[OII]$ lines in order to remove late-type
galaxies and contains 41 field galaxies and 113 in clusters.

For these systems, circularized HST I-band effective radii, \Re, and
S\'ersic fitting indices, $n$, are available.

The average slit width of the spectral observations, from which velocity
dispersion have been derived, has been converted
to an equivalent circular aperture of radius $\approx 1.025
\left(\delta x / \pi \right)$, where $\delta x$ is the slit width, in
arcsec. The ratio between spectral
aperture and the effective radius, $R_{\rm ap}/\Re$ amounts to
$\sim 1.5$, with a tail to higher ratios.

As reference stellar masses, we have taken the ones from
\cite{Saglia+10}, using rest-frame absolute photometry derived
from SED fitting (\citealt{Rudnick+09}), adopting the calibrations
of \cite{Bell_deJong01}, with a "diet" Salpeter IMF and $B - V$
colors. These masses are re-normalized to a Chabrier IMF
subtracting $\sim 0.1$ dex, accordingly to results in
\cite{Tortora+09}. Our masses are in very good agreement with those
in \cite{Vulcani+11}.

\subsection{High $z$ sample: dynamics of EDisCS galaxies}\label{sec:dyn_analysis}

Following the analysis in \cite{Tortora+09} and \cite{SPIDER-VI}
we model the velocity dispersion of each individual galaxy using
the spherical isotropic Jeans equations and hence estimate the
(total) dynamical mass \Mdyn\ (which, hereafter, we will refer to
as \Mtot) within $r=$~1~\Re.

In the Jeans equations, the stellar density is provided by the
S\'ersic fit of the photometric data, and the total (DM + stars)
mass is assumed to have the form of a Singular Isothermal Sphere
(SIS), from which $M(r)\propto \sigma_{\rm  SIS}^{2}  r$, where
$\sigma_{\rm  SIS}$ is the model (3D) velocity dispersion.

The isothermal profile has been found to provide a robust
description of the mass distribution in massive ETGs (e.g.,
\citealt{Kochanek91}; \citealt{Bolton+06_SLACSI};
\citealt{Koopmans+06_SLACSIII}; \citealt{Gavazzi+07_SLACSIV};
\citealt{Bolton+08_SLACSV}; \citealt{Auger+09_SLACSIX};
\citealt{Auger+10_SLACSX}; \citealt{Chae+14}; \citealt{Oguri+14}),
and in particular the massive dark halos dominating the outer
regions of galaxies (\citealt{Benson+00,MH02};
\citealt{Napolitano+05}; \citealt{Gavazzi+07_SLACSIV};
\citealt{vdB+07}). This ``conspiracy'' (\citealt{Rusin+03, TK04,
Koopmans+06_SLACSIII, Gavazzi+07_SLACSIV};
\citealt{Auger+10_SLACSX}) seems to be motivated also by
theoretical arguments as the stellar body and dark halos can
produce an overall isothermal profile after having gone through
the processes of gas contraction and star formation
(\citealt{Koopmans+06_SLACSIII}; \citealt{Remus+13}). For further
details on the systematics introduced by the particular model
choice, one can refer to \cite{Tortora+09} and \cite{SPIDER-VI}
(see also \citealt{Cardone+09,CT10,Cardone+11SIM}). However, in
Sect. \ref{sec:systematics} we will discuss the impact on our
results of the slope of the galaxy model adopted.

\subsection{Reference $z\sim0$ samples}

We use three local data-sets of ETGs to compare with high$-z$
results. For these samples we have performed the same dynamical analysis
and use similar assumptions for the stellar and total density
profiles as the ones for the high$-z$ sample discussed in
Sect.~\ref{sec:dyn_analysis}. Here below, some details of the local data-sets.

\begin{itemize}

\item {\it SPIDER.} This is the widest local data sample analyzed
in this paper, consisting of $\sim 4300$ giant ETGs in the
redshift range of $z=0.05$--$0.1$ (\citealt{SPIDER-I}). The
data-set includes optical$+$near-infrared photometry [from the
Sloan Digital Sky Survey (SDSS)  and the UKIRT Infrared Deep Sky
Survey-Large Area Survey], high-quality measurements of galactic
structural parameters (effective radius \Re\ and S\'ersic index
$n$), and SDSS central-aperture velocity dispersions $\sigAp$. The
sample galaxies are defined as bulge dominated systems with
passive spectra while late-type systems are removed through the
SDSS classification parameters based on the spectral type and the
fraction of light which is better described by a \cite{deVauc48}
profile (see T+12 for further details). The sample is $95\%$
complete at the stellar mass $\mst = 3 \times 10^{10}\, \rm
\Msun$, which corresponds to $\sigAp \sim 160 \, \rm km/s$. The
SPS-based stellar mass-to-light ratios, \Yst, were derived by
fitting \citet{BC03} models to the multi-band photometry, assuming
a Chabrier IMF (see \cite{SPIDER-V} and \cite{SPIDER-VI} for
further details). A de-projected S\'ersic law in the $K$-band is
used to describe the density profile of the stellar component.
Then, both the light profile and total masses are calculated at
g-band \Re, which rest-frame is approximately the observed I-band
from EDisCS. The trends at $R_{\rm e, K}$ will be discussed too.

\item {\it T+09}. This data-set contains optical photometry and
kinematics of the galaxy central regions from \cite{PS96}. The
colours are measured within 1~\Re, the central velocity
dispersions \sigc\ are measured in circular aperture of radius
\Re/8 from long slit spectra. Non-homology in the light profile is
taken into account using the $n-L_{\rm B}$ correlation in
\cite{PS97}. \vmax\ is defined as the quadratic sum of the maximum
rotation on the major and minor axes and is taken into account in
the dynamical modelling. Selecting galaxies with at least two
measured colours, brighter than $M_{B}=-16$ and further limiting
the sample to ellipticals and lenticulars we are left with $\sim
360$. See \cite{Tortora+09} for further details on the sample
selection and analysis. B-band, as g-band, approximates very well
the rest-frame of I-band measurements in EDisCS galaxies.

\item {\it \atlas3d.} This sample is originally constituted of 260
ETGs from the \atlas3d\ survey (\citealt{Cappellari+13_ATLAS3D_XV,
Cappellari+13_ATLAS3D_XX}). About 15\% of the sample have
significant gradients of the stellar mass-to-light ratio, as
inferred by their young stellar populations (H$\beta$ equivalent
width greater than 2.3~\AA). We exclude these systems and retain a
sample of 224 galaxies. The relevant data for each galaxy include
the effective radius, \Re\ (obtained by renormalizing their \Re\
from MGE light profiles with a factor of 1.35), the projected
stellar velocity dispersion, \sige, within a circularized aperture
of radius \Re, the $r$-band total luminosity $L_r$ and stellar
\ML\ ($\Upsilon_*$) derived by SPS fitting of the spectra with
\cite{Vazdekis+12} models and a \cite{Salpeter55} IMF. The
\cite{Chabrier01} IMF yields stellar masses that are $\sim 0.26$
dex smaller (see also \cite{Tortora+14_MOND} for further details).
\end{itemize}

These three local samples allow us to evaluate the impact on the
results of: a) the prescription for light distribution profiles,
b) the apertures used to measure velocity dispersion and c) SPS
prescriptions for stellar mass estimates. As we will see, the
trends are nearly independent of the sample adopted, making the
results quite robust.

\begin{figure*}
\centering\psfig{file=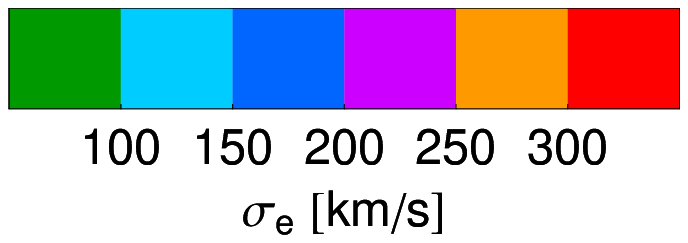, width=0.45\textwidth}
\psfig{file=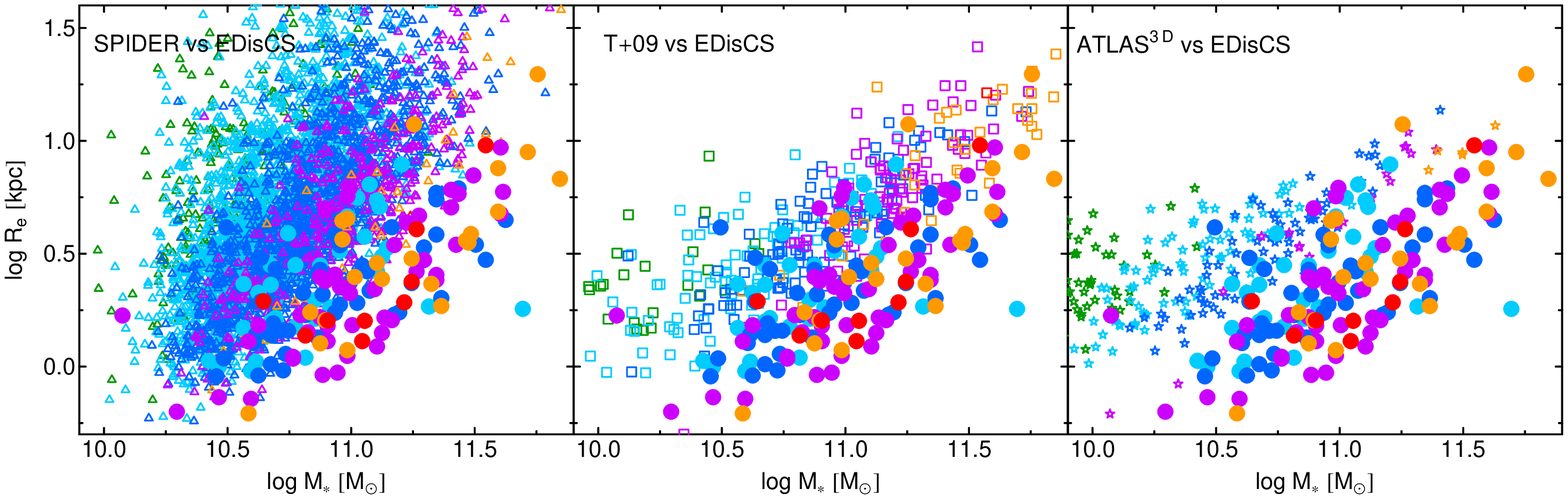, width=0.9\textwidth} \caption{Size-mass
relation colorised according to the \sige\ bins in the colour bar
in the top. We show single data-points for single galaxies: I-band
EDisCS (points), g-band SPIDER (open triangles), B-band T+09 (open
squares) and r-band \atlas3d\ (open
stars).}\label{fig:size-sigma-mass_2}
\end{figure*}

\begin{figure*}
\centering \psfig{file=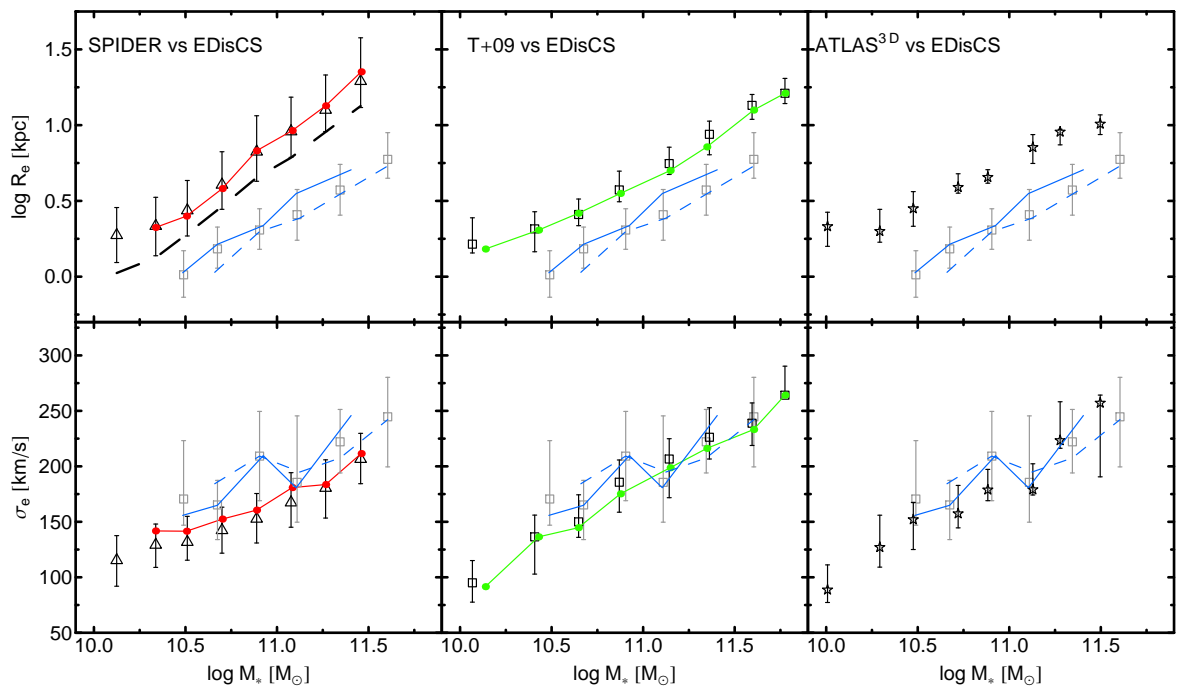, width=0.9\textwidth}
\caption{Size-mass (top panels) and \sige-mass (bottom panels)
relations. Medians and 25-75th quantiles are plotted. EDisCS
sample is plotted as gray symbols, I-band \Re\ are used. The
average trends for lower-z ($z\leq 0.6$, blue lines) and
higher-$z$ ($z>0.6$, blue dashed lines) EDisCS galaxies are also
plotted. See text for further details about bin selection. From
left to right we compare EDisCS results with g-band (open
triangles) and K-band (dashed line) from SPIDER, B-band from T+09
(open squares) and r-band from \atlas3d\ survey (open stars). Red
(green) points and lines are for results corrected for progenitor
bias for SPIDER (T+09) data-sets.}\label{fig:size-sigma-mass}
\end{figure*}

\begin{table*}
\centering \caption{Best-fit parameters for the relation $\log \Re
= a + b \log \left( \mst/10^{11} \right)$ and $\log \sige = a + b
\log \left(\mst/10^{11} \right)$ for the samples analyzed. Best
value, 1 $\sigma$ error and significance of the correlation are
shown.}\label{tab:tab1}
\begin{tabular}{lcccccc} \hline
& \multicolumn{3}{c}{\Re-\mst} & \multicolumn{3}{c}{\sige-\mst}\\
\hline
\rm Sample & $a$ & $b$ & Sig & $a$ & $b$ & Sig \\
\hline
SPIDER & $0.89\pm 0.01$ & $0.81\pm 0.04$ & 99\% & $2.212 \pm 0.004$ & $0.18 \pm 0.02$ & 99\% \\
T+09 & $0.69\pm 0.01$ & $0.58\pm 0.03$ & 99\% & $2.265\pm 0.007$ & $0.27\pm 0.01$ & 99\% \\
\atlas3d\ & $0.75\pm 0.02$ & $0.50\pm 0.04$ & 99\% & $2.28\pm 0.01$ & $0.29\pm 0.02$ & 99\% \\
\hline
EDisCS & $0.35\pm 0.03$ & $0.61\pm 0.08$ & 99\% & $2.29\pm 0.02$ & $0.15\pm 0.04$ & 99\% \\
EDisCS -- low-$z$ & $0.42\pm 0.05$ & $0.69\pm 0.12$ & 99\% & $2.29\pm 0.03$ & $0.19\pm 0.06$ & 99\%\\
EDisCS -- high-$z$ & $0.31\pm 0.04$ & $0.66\pm 0.10$ & 99\% & $2.3\pm 0.02$ & $0.11\pm 0.06$ & 99\%\\
\hline
\end{tabular}
\end{table*}

\section{Size-mass and Faber-Jackson evolution}\label{sec:correlations}

A simple monolithic-like scenario, where the bulk of the stars is
formed in a single dissipative event followed by a passive
evolution, is not longer supported by the observations, while it
is becoming increasingly evident the occurrence of a strong mass
and size evolution in ETGs (\citealt{Trujillo+06};
\citealt{Trujillo+11}).

For the analysis we want to perform, we are interested at
comparing some relevant correlations found in local galaxies, as
the ones between the galaxy size, i.e. \Re, and the stellar mass,
\mst, and the  one between the velocity dispersion and \mst\ (see,
e.g., \citealt{Tortora+09}; \citealt{SPIDER-I};
\citealt{Cappellari+13_ATLAS3D_XV}) with the same for high$-z$
galaxies (\citealt{Saglia+10}).

In Fig. \ref{fig:size-sigma-mass_2}, we show the \Re--\mst\
relation for each galaxy of the four samples colorized according
to their central velocity dispersion. This figure provides
insights on the covariance among the stellar and dynamical
parameters relevant in our analysis; we will adopt a similar
approach in the next Section when we will discuss the central DM
content. Both low- and high-$z$ galaxies follow the typical
positive \Re--\mst\ relation. It is also evident that a similar
positive correlation exists between \sige\ and \mst, independently
of the redshift, as higher-\sige\ galaxies are more concentrated
at the higher stellar masses and lower-\sige\ are found toward
lower stellar masses both for high-$z$ and the low-$z$ samples.
This is made clearer in Fig.~\ref{fig:size-sigma-mass}, where we
show the median results of the EDisCS sample compared face-to-face
to the three local samples. The correlations for each data sample
are quantified with a log--log fit, and the best-fit parameters
are reported in Table~\ref{tab:tab1}.

Consistently with the local observations and independently of
differences in the bands adopted for the \Re s, the EDisCS sample
presents the same size-mass trend, with more massive galaxies
being lerger (top panels in Fig.~\ref{fig:size-sigma-mass} and
Table~\ref{tab:tab1}). All the correlations are significant at
more than $99\%$. Moreover, at fixed mass, the high$-z$ galaxies
in the EDisCS sample show a systematic offset toward smaller radii
with respect to the same quantities for the local samples, i.e.
they are more compact at all masses. The same is found for the
S\'ersic indices (not shown) which are smaller than the local
values. Instead, the slope of the size-mass for the EDisCS sample
is identical to the ones of most of the samples analyzed and a bit
shallower than SPIDER sample\footnote{The shallower size-mass
relations found in T+09 and \atlas3d\ with respect to SPIDER
sample are explained by the fact that fitting a high-n galaxy with
a de Vaucouleurs profile gives a systematically smaller \Re\
value, hence flattening the size-mass relation.}.  The high-$z$
galaxies are, on average, $\sim 2-4$ times smaller than local
SPIDER galaxies while only $\sim 2$ smaller than the other local
samples (T+09 and \atlas3d). If K-band \Re s are used for SPIDER
galaxies, then the difference with respect to the high-$z$ sample
becomes more similar to that found for the other local samples.

We have also split the high$-z$ sample in two further bins to see
whether there is the signature of a further evolution with z
within the EDiSC sample itself. The two subsamples are selected to
have redshifts $\leq 0.6$ and $>0.6$, and have median redshifts of
$0.52$ and $0.75$, respectively.  We find that the trends are
almost unchanged (see Table \ref{tab:tab1}) and that there is
possibly a further evolution in the intrinsic size toward more
compact \Re\ at higher$-z$ (\citealt{Saglia+10}).

Going to the correlations between velocity dispersion and stellar
mass, this is also seen in the EDisCS sample (bottom panels in
Fig.~\ref{fig:size-sigma-mass} and Table~\ref{tab:tab1}). At fixed
stellar mass, no significant evidence of evolution is found with
respect to the local samples, except a marginal offset at $\mst
\lsim 10^{11}\, \rm \Msun$, which shows smaller \sige\ than the ones in
EDisCS. In this case the spitting of the EDisCS in two $z$ bins does
not show any further evolution signatures (see Fig.
\ref{fig:size-sigma-mass} and Table \ref{tab:tab1}).

This mild evolution in velocity dispersion is consistent with
predictions from galaxy mergers (\citealt{Hopkins+09_DELGN_IV};
\citealt{Cenarro_Trujillo09}), but not with the puffing up
scenario in \cite{Fan+08}, which would produce a stronger change
with redshift.

\begin{figure*}
\centering \psfig{file=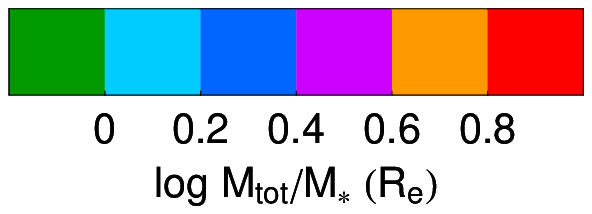,width=0.45\textwidth}
\psfig{file=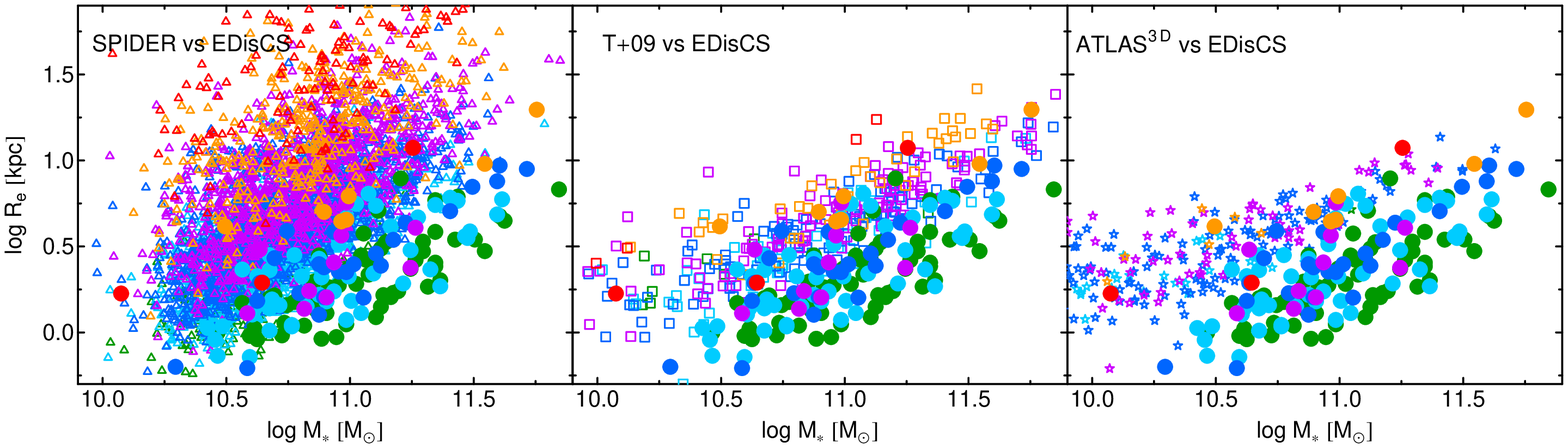,width=0.9\textwidth}
\psfig{file=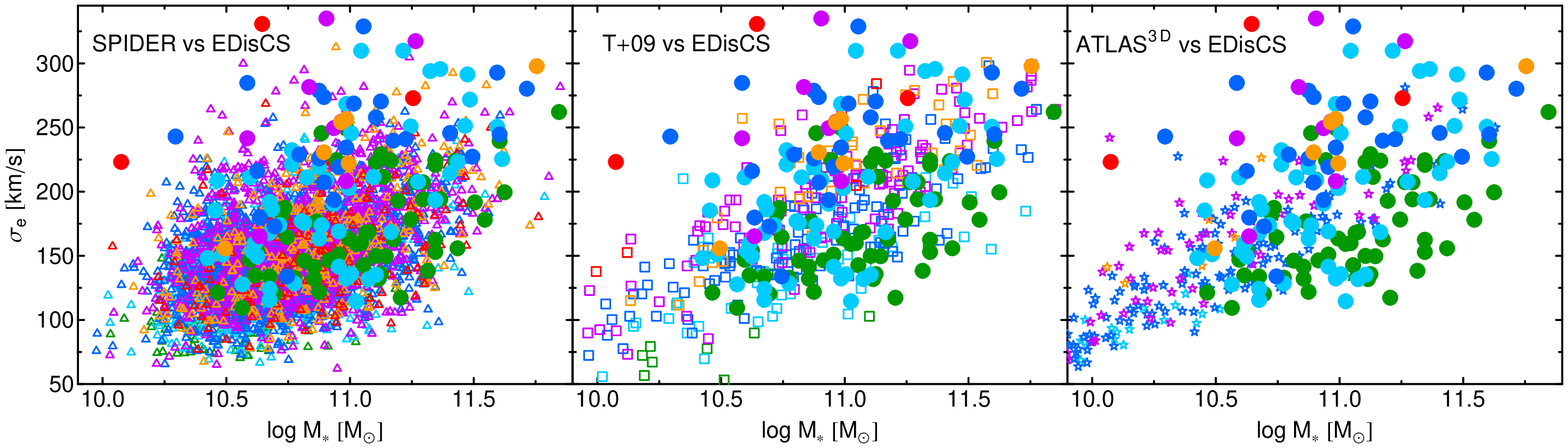,width=0.9\textwidth} \caption{Size-mass (top
panels) and \sige-mass (bottom panels) relations colorised
according to the \TtoSM\ bins in the colour bar in the top. We
show single data-points as in Fig.
\ref{fig:size-sigma-mass_2}.}\label{fig:size-mass-fdm}
\end{figure*}

The results shown in Figs. \ref{fig:size-sigma-mass_2} and
\ref{fig:size-sigma-mass} do not take into account the progenitor
bias, for which low-$z$ ETG samples contain galaxies that have
stopped their star formation only recently and that would not be
recognized as ETGs at higher redshifts. In particular, a
morphologically selected local sample of ETGs as the ones we have
analyzed, contains systems with relatively young ages that, when
evolved back to match the high-$z$ sample, would not be recognized
as being passive objects (\citealt{vanDokkum_Franx01};
\citealt{Saglia+10}; \citealt{Valentinuzzi+10_WINGS,
Valentinuzzi+10_EDisCS}). For the SPIDER sample,
luminosity-weighted ages are available (\citealt{SPIDER-V}). These
ages are derived from the spectral fitting code STARLIGHT
(\citealt{CidFernandes+05}), which finds the best combination of
SSPs (with free age, metallicity and alpha-enhancement) which
reproduce the measured spectra. Thus, we only take those SPIDER
objects which have an age $\geq \, 1.5, \rm Gyr$, at the average
redshift of the EDisCS sample (i.e., older than 7.5 Gyr). On the
contrary, for the T+09 data sample, ages estimated fitting
exponential SFs (with characteristic timescale $\tau$) to optical
colours, are available (\citealt{Tortora+09}; \citealt{NRT10}).
For these reasons we have only taken those systems that after a
time $\tau$ (corresponding to the epoch when the SF is reduced to
$37\%$ of the initial value) are older than the look-back time at
the average redshift of the sample. The results are fairly
unchanged if we use the epoch when the SF is $\sim 14\%$ of the
initial value (i.e. at the epoch $2 \tau$). Ages for \atlas3d\ are
not available, for this reason the impact of progenitor bias will
not be discussed for this dataset.

The outcomes for such old systems are shown in Fig.
\ref{fig:size-sigma-mass} as red and green lines for SPIDER and
T+09 data samples, respectively. The impact on the results is very
weak (e.g., \citealt{BNE14}), and it is possibly more relevant in
the \sige--\mst\ correlation. We notice that in the two
datasamples the impact of the progenitor bias pushes the \Re\ in
the two opposite directions, since the correlation between \Re\
with galaxy age is still controversial. In fact, contrasting
results are found by observational analysis, which find that at
fixed mass/$\sigma$ younger systems are larger
(\citealt{Shankar_Bernardi09}; \citealt{NRT10};
\citealt{Valentinuzzi+10_WINGS}) or are as sized as older galaxies
(\citealt{Graves+09}). The outcomes from semi-analytic galaxy
formation models are also still unclear, as there are results
showing that younger galaxies are larger
(\citealt{Khochfar_Silk06}) or also smaller (\citealt{Shankar+10})
than the oldest systems.

Despite these uncertainties, in the following we will discuss
the results with and without the progenitor bias for
those samples for which this has been computed (i.e. SPIDER and T+09).

\section{Dark matter evolution}\label{sec:evolution}

\begin{figure*}
\centering \psfig{file=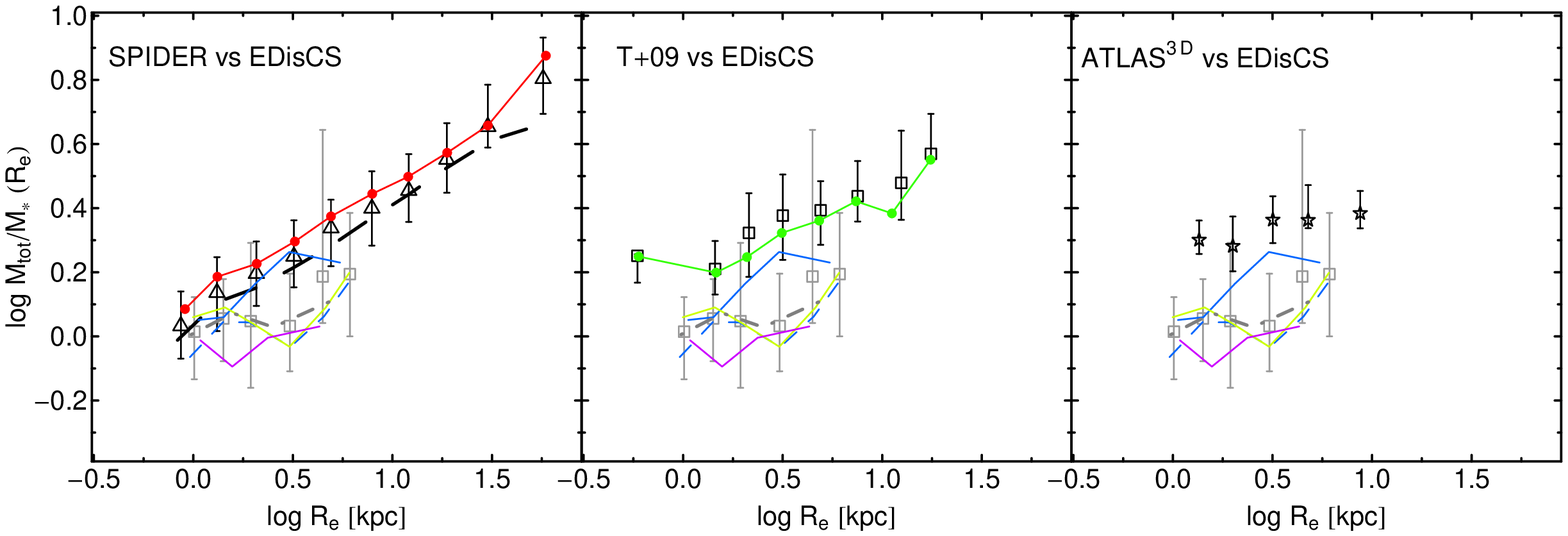,width=0.9\textwidth}
\psfig{file=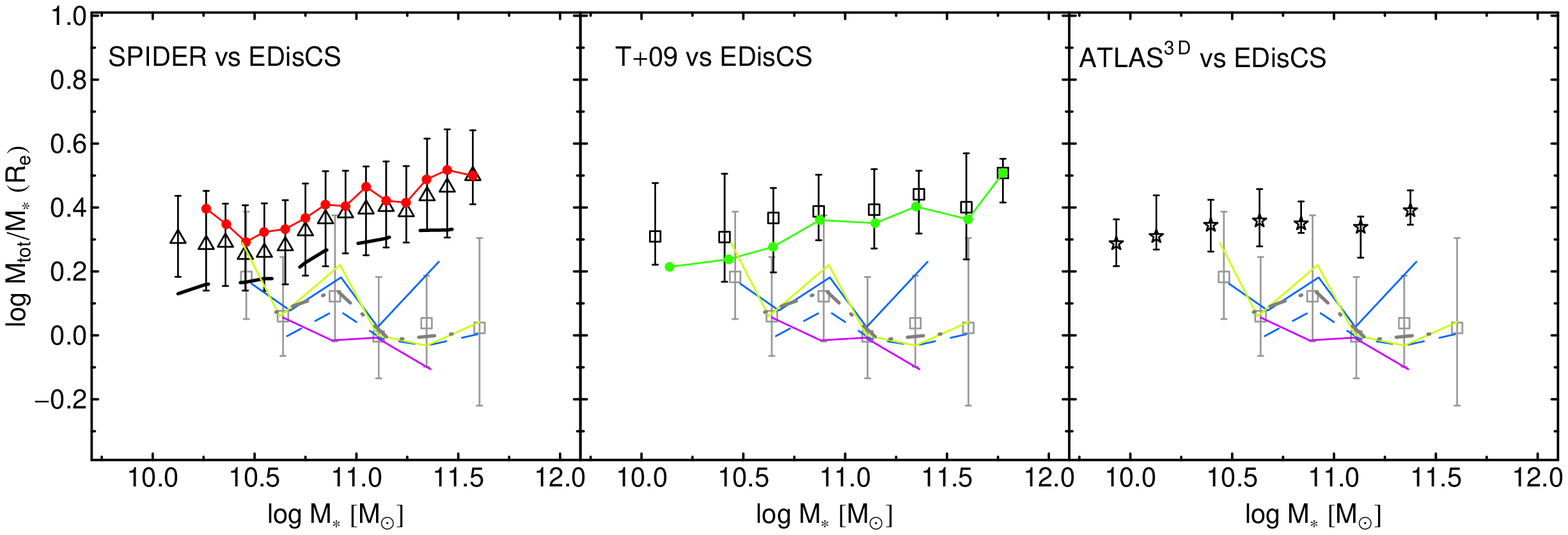,width=0.9\textwidth}
\psfig{file=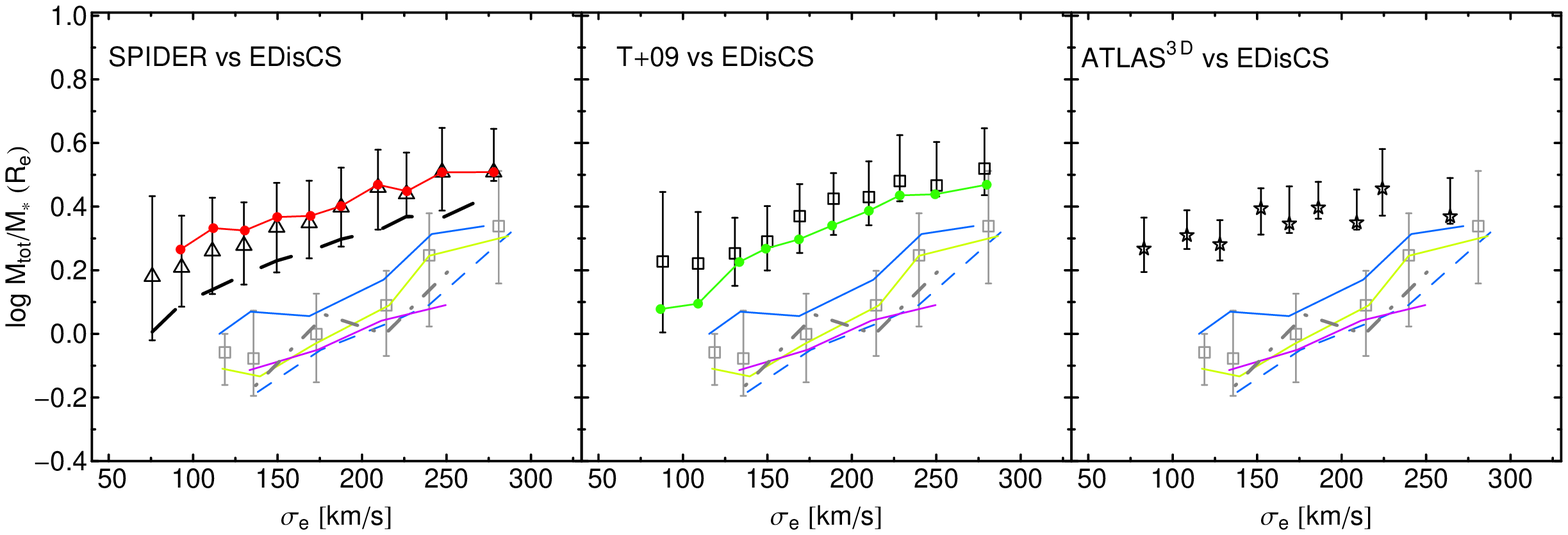,width=0.9\textwidth} \caption{\TtoSM\ (in
logarithmic scale) as a function of \Re\ (top panels), \mst\
(middle panels) and \sige\ (bottom panels). Symbols are as in
Fig.~\ref{fig:size-sigma-mass}. The dashed line is made
calculating the SPIDER K-band profiles at $R_{\rm e,K}$. The
yellow lines are for the roundest EDisCS objects with axial ratio,
$q>0.6$. Purple lines are for field EDisCS systems. Gray
point-dashed lines are for EDisCS elliptical morphologically
selected (i.e. with $T \geq -4$).}\label{fig:fdm}
\end{figure*}

In this section we investigate the galaxy central DM content as a
function of $z$. To quantify that, we will use either the
de-projected DM fraction at \Re, $\fdm = 1 - \mst(\Re) /
\Mtot(\Re)$, or the deprojected total-to-stellar mass ratio at
\Re, $\Mtot(\Re)/\mst(\Re)$. This latter will be useful to avoid
unphysical negative \fdm\ values, in particular when computing the
best-fit relations among galaxy parameters.

From the previous section we have seen that there is almost no
evolution in the relation \sige$-\mst$ and this might suggest that
so would be for the $\Mtot(r)/\mst(r)$, if the \sige\ is a fair
proxy of the total mass. On the other hand, we have observed a
strong evolution of the \Re\ with the stellar mass, which shows
that the scale where most of the mass in stars is confined was
more compact in the earlier epochs with respect to present time
(even in case the progenitor bias is taken into account). Thus, in
the second part of the present section we will discuss the role of
this size evolution on our DM inferences.

\begin{table*}
\centering \caption{Best-fit parameters for the relation \TtoSM --
\Re, \TtoSM -- \mst\ and \TtoSM -- \sige. The following best
relations are fitted: $\log \TtoSM = a + b \log \left( \Re/2\rm
kpc \right)$, $\log \TtoSM = a + b \log \left( \mst/10^{11}\Msun
\right)$ and $\log \TtoSM = a + b \log \left( \sige/200\rm km/s
\right)$. Best value, 1 $\sigma$ error and significance of the
correlation are shown.}\label{tab:tab2}
\begin{tabular}{lccccccccc}
\hline
& \multicolumn{3}{c}{\TtoSM -- \Re} & \multicolumn{3}{c}{\TtoSM -- \mst} & \multicolumn{3}{c}{\TtoSM -- \sige}\\
\hline
\rm Sample & $a$ & $b$  & Sig & $a$ & $b$ & Sig & $a$ & $b$ & Sig \\
\hline
SPIDER & $0.18\pm 0.01$ & $0.40\pm 0.02$ & 99\% & $0.38 \pm 0.01$ & $0.17 \pm 0.03$ & 99\% & $0.05 \pm 0.03$ & $0.36 \pm 0.03$ & 99\% \\
T+09 & $0.29\pm 0.02$ & $0.28\pm 0.04$ & 99\% & $0.40\pm 0.02$ & $0.06\pm 0.04$  & 99\% & $0.05 \pm 0.06$ & $0.36 \pm 0.06$ & 99\% \\
\atlas3d\ & $0.31\pm 0.02$ & $0.16 \pm 0.05$ & 99\% & $0.36\pm0.01$ & $0.04\pm0.02$ & 99\% & $0.21\pm0.05$ & $0.17\pm0.06$ & 99\%\\
\hline
EDisCS & $0.07\pm 0.03$ & $0.20\pm 0.10$ & 99\% & $0.08\pm 0.03$ & $-0.17\pm 0.12$ & 95\% & $-0.42 \pm 0.13$ & $0.52 \pm 0.12$ & 99\% \\
EDisCS -- low-$z$ & $0.14\pm 0.05$ & $0.26\pm 0.13$ & 99\% &
$0.16\pm 0.05$ & $0.05\pm 0.16$ & 95\% & $-0.27 \pm 0.10$ & $0.46
\pm 0.11$ & 99\% \\
EDisCS -- high-$z$ & $0.01\pm 0.05$ & $0.23\pm 0.15$ & 99\% & $0.03\pm 0.05$ & $-0.08\pm 0.20$ & 95\% & $-0.59 \pm 0.12$ & $0.61 \pm 0.12$ & 99\% \\
\hline
\end{tabular}
\end{table*}

\subsection{DM content and correlations}\label{sec:DM_corr}

In Fig. \ref{fig:size-mass-fdm} we start by re-proposing the
\Re--\mst\ and \sige--\mst\ correlations (see Fig.
\ref{fig:size-sigma-mass_2}), now colorised according to the
\TtoSM. The average \TtoSM\ are also plotted as a function of \Re,
\mst\ and \sige\ in Fig. \ref{fig:fdm} and the best-fitted
relations listed in Table \ref{tab:tab2}.

\subsubsection{Results}

As for the \Re--\mst\ and \sige--\mst\ correlations, in local
galaxies the DM trends are nearly independent of the sample
adopted. In agreement with the local samples, the high-$z$
galaxies from EDisCS preserve almost all the DM trends against
stellar mass, central velocity dispersion and effective radius
(see Figs. \ref{fig:size-mass-fdm} and \ref{fig:fdm}), with almost
all correlations being significant at more than $99\%$. However, a
clear offset in the correlations is evident. In Fig.
\ref{fig:size-mass-fdm} we find that at fixed \mst, EDisCS
galaxies present similar \sige, but smaller \Re\ and \TtoSM\ with
respect to local galaxy samples. This is made even clearer in Fig.
\ref{fig:fdm} where the \TtoSM\ median relations are shown. Here
the EDisCS galaxies are more DM dominated at larger \Re\ and
\sige\ (i.e. they follow a positive correlation) along a trend
which runs almost parallel to the local sample but shifted toward
lower \TtoSM\ (see Table \ref{tab:tab2}). In particular, we note
that at any fixed \Re, the EDisCS sample presents \TtoSM\ of $\sim
0.15-0.25$ dex smaller than the ones at low$-z$. These differences
are all significant at more than $3\sigma$ as evident from the
linear regressions in Table \ref{tab:tab2}. If \Re\ was the only
responsible of the $\TtoSM-z$ variation one would expect
statistically no variation of the $\TtoSM-\Re$ with $z$.

In the same Fig. \ref{fig:fdm}, we also observe that the change in
\TtoSM\ is stronger at any fixed \mst\ and \sige, being the
average offset between the high and low redshift samples of
\TtoSM\ $\sim 0.3-0.35$ dex and $\sim 0.5-0.6$ dex, respectively
for the two quantities. Particularly interesting is the
$\TtoSM-\mst$ correlation which changes the slope from the low to
the high$-z$ trend, although consistent within the errors with
null value (see Table \ref{tab:tab2}). This latter, together with
the presence of a marginal evolution also of the $\TtoSM-\Re$ in
the top panel, show that the \TtoSM\ evolution with $z$ cannot be
driven by the \Re\ only, although previous analyses have shown
that the main driver of most of the scaling relations involving
the central DM fraction in ETGs is the effective radius (e.g.,
\citealt{NRT10}; \citealt{Auger+10_SLACSX};
\citealt{Tortora+10lensing, SPIDER-VI}) and there are accumulating
evidences that \Re\ scales with the redshift
(\citealt{Trujillo+06}; \citealt{Saglia+10};
\citealt{Trujillo+11}, see also the top panels of Fig.
\ref{fig:size-sigma-mass}). Despite the fact that the \Re$-z$
relation might be a key factor for interpreting the $\TtoSM-z$
trend, there have to be mechanisms related to the global stellar
and dark mass assembly which also shape the \TtoSM\ trend with
$z$. We will discuss the possible the evolution scenarios later in
\S\ref{sec:evol}, here we note that, taking the current \TtoSM\
results at face values, it is evident from Figs.
\ref{fig:size-mass-fdm} and \ref{fig:fdm} the evolution of this
quantity, which has increased with time across the last $\sim 7 \,
\rm Gyr$. Indeed, this observed evolution in \TtoSM\ is also found
if we split EDisCS sample in two redshift bins (see Fig.
\ref{fig:fdm} and Table \ref{tab:tab2}).

\begin{figure}
\centering
\psfig{file=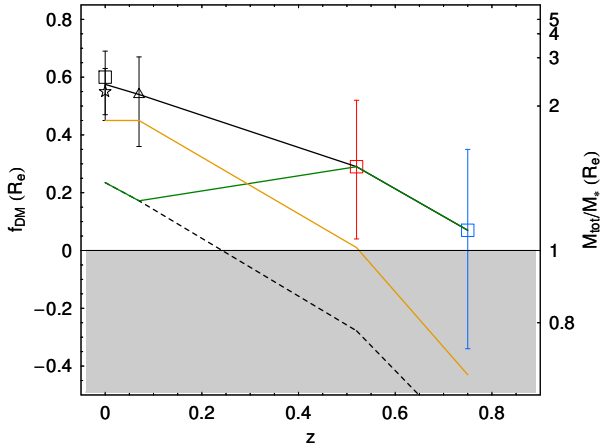,width=0.45\textwidth}\caption{Average DM
fraction evolution. Symbols for local samples, assuming a Chabrier
IMF are as in Figs. \ref{fig:size-sigma-mass} and \ref{fig:fdm}.
Red and blue symbols are for lower-z and higher-z EDisCS galaxies.
Solid line connect these datapoints. Dashed line connects the
results for a Salpeter IMF. The gray region set the locus of
unphysical (i.e. negative) DM fractions. Solid green line is for
toy-model assuming a Chabrier (Salpeter) IMF at high (low)
redshift. Solid orange line is for a toy-model assuming the local
\dimf--\sige\ relation from
\citet{TRN13_SPIDER_IMF}.}\label{fig:final}
\end{figure}

In Fig. \ref{fig:final} we summarise the average results for the
Chabrier IMF and the bottom-heavier Salpeter IMF, this time
represented as the DM fraction, \fdm, as a function of redshift
(on the right axes the corresponding \TtoSM\ are also shown). It
is now explicitly evident the increasing trend of the \fdm\ at
lower $z$, which is the consequence of the offset of the \TtoSM\
between the low- and high-$z$ samples, as discussed before. In the
same figure we plot the regions of ``unphysical'' \fdm\ as a
shadowed area. Here the lower \TtoSM\ of the high$-z$ sample
(green points in Fig. \ref{fig:size-mass-fdm}) implies a much
higher fraction of (unphysical) negative \fdm\ (57 out 154
galaxies, i.e. 37\%) assuming a Chabrier IMF, which becomes even
more dramatic if a Salpeter IMF is assumed (115 out of 154, i.e.
75\%). A fraction of these galaxies with negative \fdm\ would be
compatible with observational scatter in \mst\ and \Mtot\ (e.g.
Appendix A in \citealt{NRT10}). However, it is difficult to
explain all the negative \fdm\ as a consequence of this
observational scatter, rather this may suggests that the
bottom-heavy Salpeter IMF is disfavoured with respect to a
Chabrier IMF.

Hence, the inferred \fdm\ from a Salpeter IMF (or bottom-heavier
IMFs in general) are disfavoured with respect to a Chabrier IMF,
mainly at high redshift ($z \gsim \, 0.6$). In order to reduce
this tension, one can relax the assumption of IMF universality,
and re-compute the effect on the median \fdm s as a function of
$z$ by assuming a Salpeter and Chabrier IMFs in local and EDisCS
samples, respectively: this very crude assumption would cancel out
the variation of \fdm\ with redshift as shown by the green line in
Fig. \ref{fig:final}.

A more reasonable assumption for a non-universal IMF should take
into account also some variation in terms of mass, as found in
local analysis (\citealt{Conroy_vanDokkum12b};
\citealt{Cappellari+12}; \citealt{Spiniello+12};
\citealt{Dutton+13}; \citealt{Ferreras+13};
\citealt{Goudfrooij_Kruijssen13};
\citealt{LaBarbera+13_SPIDERVIII_IMF}; \citealt{TRN13_SPIDER_IMF};
\citealt{Weidner+13_giant_ell}; \citealt{Tortora+14_MOND}). We
have evaluated also this effect by correcting the DM fractions for
all the samples using the same local relation between $\dimf =\Yst
/ \Upsilon_{\star, Chabrier}$ and \sige\ from
\cite{TRN13_SPIDER_IMF}. In this case some \fdm\ trend with
redshift is still observed (see orange line in Fig.
\ref{fig:final}), but the number of high-$z$ galaxies with
negative DM (mainly with small \sige) is still too large to
accommodate these systems without a variation of IMF across the
cosmic time.

{\it The net conclusion of this analysis is that according to the
\fdm\ variation with $z$, the IMF which would be compatible with
physical \fdm\ values can range from bottom-heavy to bottom-light
at low$-z$, but it does not have the same leverage at higher$-z$
(e.g. $z\gsim0.6$) where the IMF has to be mainly bottom-light
(e.g. Chabrier-like), and compatible with a Salpeter IMF in very
high-\sige\ systems only, producing negative DM fractions just in
a tiny handful of systems.}

A detailed investigation of the IMF variation with \sige\ as a
function of $z$ is beyond the scope of this paper. Therefore, in
the following we will discuss the results based on a constant IMF
across the lookback time.

\subsubsection{Systematics}\label{sec:systematics}

We have seen that the IMF is a major source of uncertainty in the
\fdm\ trend with $z$. However, there are other sources of
systematics that may provide alternative explanations for the
negative \fdm\ values of the EDisCS sample within 1 \Re\ and
eventually modify our trends with redshift. Besides systematics in
the stellar mass estimates, the negative \fdm\ values can be due
to measurement errors on galaxy parameters and/or a failure of the
mass model adopted for the \Mdyn\ (e.g. \citealt{SPIDER-VI}). A
brief discussion of all possible systematics is given here below.

\begin{description}
\item[(i)] We measure \Re s using I-band photometry, which means that \Re s are approximately rest-frame
V and B bands, respectively, at $z\sim 0.4$ and $z \sim 0.8$. \Re
s are found to be larger at smaller wavelengths
(\citealt{Sparks_Jorgensen93}; \citealt{LaBarbera_deCarvalho09};
\citealt{Roche+10}; \citealt{Vulcani+14}). In particular, using a
sample of galaxies with $z < 0.3$, \cite{Vulcani+14} estimate an
increase from g to u and from r to g band of $\lsim 15\%$ of and
similar results are found in \cite{LaBarbera_deCarvalho09}
following the method in \cite{Sparks_Jorgensen93}. For this
reason, our \Re s which are measured in the rest-frame V-band at
the smallest EDisCS redshifts are conservative lower limits and
would be larger if calculated in the same rest-frame as the
highest redshift ETGs. Thus, stronger evolution of \Re\ and \fdm\
with redshift is expected if this effect would be taken properly
into account. We have shown the impact of waveband in Fig.
\ref{fig:fdm} for the SPIDER \TtoSM\ where the profiles are
calculated at K-band \Re, which are smaller than our reference
values. This reduces the difference with EDisCS galaxies with
respect to the results from g-band light profile and g-band \Re\
(not shown in the Figure for brevity) which would give \TtoSM\
larger than our reference estimates.
\item[(ii)] In addition to uncertainties in stellar mass estimates,
the choice of the mass profile that we have made (i.e. the SIS
model) can be inappropriate for low-\sige\ (or \mst) galaxies and
cause an excess of negative \TtoSM\ values, mainly at high-$z$
(e.g. \citealt{Sonnenfeld+13_SL2S_IV}). In particular, \TtoSM\ and
the slope of total mass density are tightly correlated, with
shallower density profiles corresponding to larger \TtoSM\
(\citealt{Humphrey_Buote10}; \citealt{Remus+13};
\citealt{Dutton_Treu14}). To quantify the impact of a free total
mass density slope, $\alpha$, we have adopted a power-law mass
density $\rho \propto r^{\alpha}$, with slope steeper and
shallower than isothermal. We use the two extreme values $\alpha =
-2.5$ and $-1.5$, which bracket most of the results in the
literature. We find that the average \TtoSM\ gets smaller (larger)
of 0.12 (0.05) dex for $\alpha = -2.5$ ($=-1.5$); in a realistic
case with a varying slope with mass, smaller changes, $<$ 0.12
dex, will be found (e.g. \citealt{Dutton_Treu14}). If $\alpha$ is
constant with time, then these corrections have to be applied to
both local and EDisCS samples, and our results are left naturally
unaffected. On the other hand, if we assume for local ETGs that
the slope is varying with mass as it is effectively found, then
the only way to totally remove any \TtoSM\ evolution is that
EDisCS galaxies have total mass density slopes very shallow
($\alpha >$ -1.5, and consequently larger \TtoSM), which is not
expected, since at high redshifts, where gas and in situ star
formation dominate the galaxies, the ETGs from cosmological
simulations have a total density slope very steep ($\alpha \sim
-3$), and mergings tend to drive the galaxy to a nearly isothermal
profile (\citealt{Remus+13}). Thus, although a varying slope with
redshift has to be further investigated, our tests showed that the
conclusions of the present paper are left nearly unchanged.
\item[(iii)] If rotation velocities would be
included in the analysis \Mdyn\ would get higher, reducing the
fraction of negative \fdm s.
\item[(iv)] As discussed in Sect. \ref{sec:correlations}, galaxies
which are star forming at the
cosmic epoch of the EDisCS galaxies have to be removed. Thus,
progenitor bias has to be applied, and we plot the results in Fig.
\ref{fig:fdm}. As for the trends of \Re\ and \sige\ as a function
of \mst, the \TtoSM\ results are only little affected. In
particular, the inclusion of the progenitor bias make the \TtoSM\
larger (smaller) for SPIDER (T+09) samples. See \cite{NRT10} for
further details about the \fdm--age correlation (see also
\citealt{Tortora+10lensing}). We notice that after the progenitor
bias is applied, the median \TtoSM\ only slightly changes, since
the largest fraction of galaxies (the oldest) is left unchanged.
\item[(v)] We have checked the effect of the ellipticity in the
mass inferences and restricted the analysis to EDisCS objects with
axis-ratio $q > 0.6$ ($60\%$ of the full sample, see yellow lines
in Fig. \ref{fig:fdm}), in order to limit to the roundest galaxies.
The overall results are practically unchanged.
\item[(vi)] The results are also almost unchanged if we only take
those objects with morphological type of pure ellipticals (i.e. $T
\leq -4$). Both tests on the ellipticity (point (v)) and
morphology above give support to the negligible impact of ordered
motion on our \Mdyn\ estimates.
\item[(vii)] Finally, we find that \TtoSM\ in field galaxies are slightly
smaller than those in cluster galaxies (consistently with the
stronger evolution found in the literature, \citealt{Saglia+10}),
but due to the scatter in the sample and uncertainties this
difference is not statistically relevant, pointing to a central DM
content which is almost independent on the environment the galaxy
lives in (accordingly to similar analysis in local environments,
\citealt{SPIDER-VI}).

\end{description}

\subsection{Passive evolution vs. accretion by merging}\label{sec:evol}

We complement the analysis performed in the previous section by
discussing our results within the two dominant evolution
scenarios: i.e passive evolution from monolithic collapse and
galaxy mergers from hierarchical model.

\subsubsection{Passive evolution}

We start from discussing a pseudo-passive evolution, assuming no
merging and compute the ageing of stellar populations in our
sample. The EDisCS galaxies, as spectroscopically selected passive
objects, have low residual SF and can be fairly approximated by a
single burst SSP, or at most a short-duration SF. We consider
stellar population synthesis (SPS) models from \cite{BC03} and
adopt an exponential SF with timescale of $\tau = 500 \, \rm Myr$
(\citealt{Tortora+09}) and different formation redshifts ($z_{\rm
f} = 1.5, 2.5, 3$). On average, for passive evolution the EDisCS
stellar mass become smaller by $0.01-0.02$ dex, due to mass loss.
This very mild evolution is not enough to produce the local
\Re--\mst\ relations and similarly the local DM correlations.
Longer SF history with $\tau > 1 \, \rm Gyr$ would increase the
\mst, increasing the discrepancies with the local observed local
samples.

\begin{figure*}
\centering \psfig{file=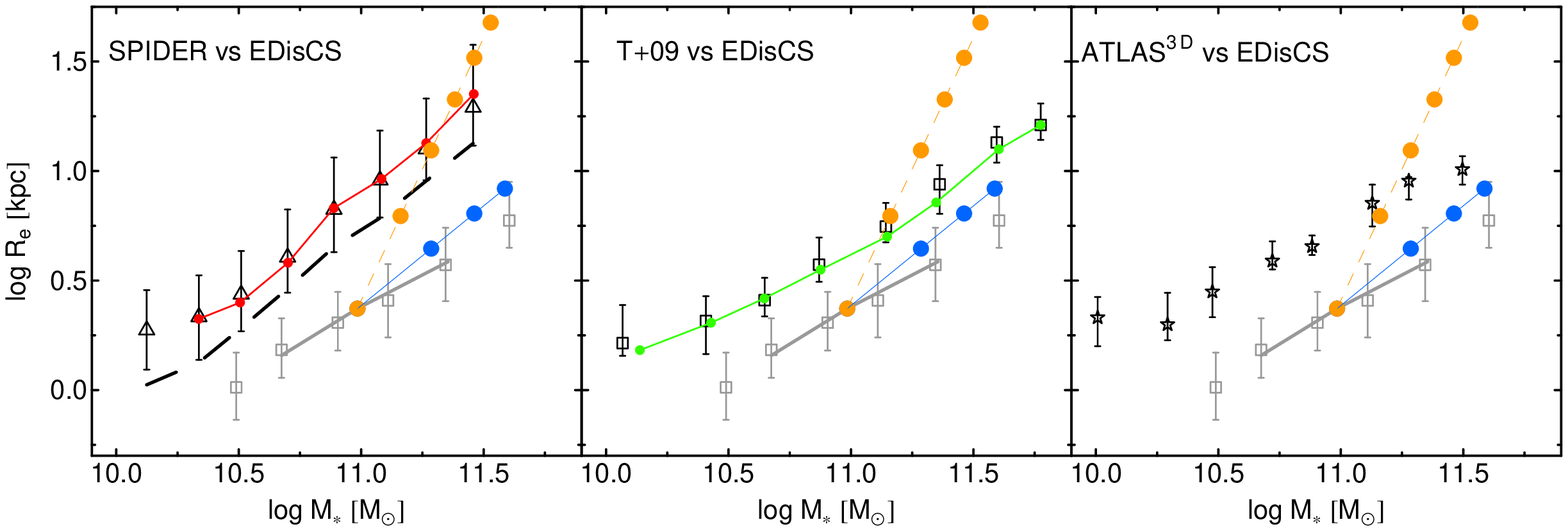,width=0.9\textwidth}
\psfig{file=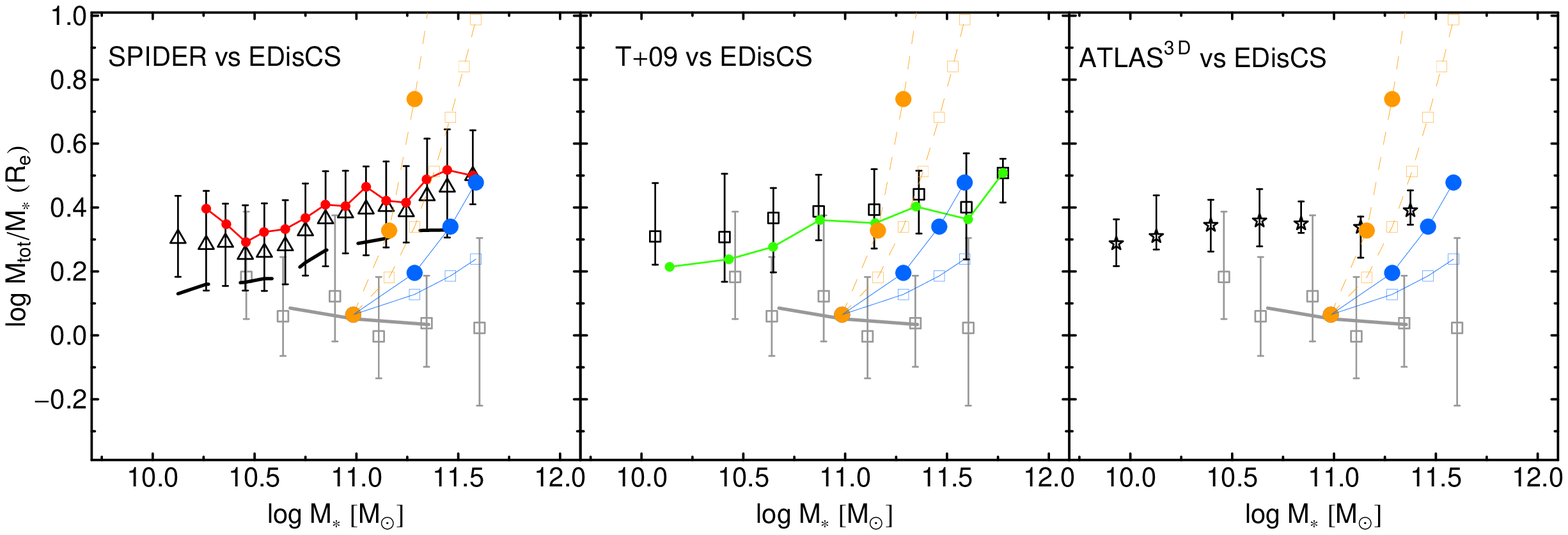,width=0.9\textwidth}\caption{Evolution of
\Re--\mst\ (top panels) and \TtoSM--\mst\ (bottom panels). The
symbols for local and EDisCS samples are the same as in Figs.
\ref{fig:size-sigma-mass} and \ref{fig:fdm}. The gray line is for
EDisCS sample, but averaging over only three mass bins. We take as
example, the average galaxy at $\log \mst/\Msun = 11$ and evolve
it accordingly to the toy-models discussed in the text and in
Table \ref{tab:tab3}. Blue and orange tracks are for major and
minor mergings, respectively. Dots and squares are for NFW and
AC+NFW profiles, and set the \Re, \mst\ and \TtoSM\ after each
single merging. Of course, the DM model only impacts the
\TtoSM--\mst\ trends. Notice that for the minor merging case, dots
and squares mark the epochs when one-half of the initial mass is
accreted.}\label{fig:evolution}
\end{figure*}

\subsubsection{Hierarchical scenario}

Alternatively, galaxy mergers seem the natural mechanisms that can
account for both size and mass accretion. Simulations of
dissipationless major mergers of elliptical galaxies in
\cite{Boylan-Kolchin+05} have predicted the change in central DM
in the merger remnant. They demonstrate that the DM fraction
within a certain physical radius decreases a bit after the merger.
But the DM fraction within the final \Re\ is greater than the DM
fraction within the initial \Re, because $M_{\rm tot}(\Re)$
changes after the merger more than $\mst(\Re)$ (see also below for
more quantitative details).

The effect of merging on the central \fdm\ has been investigated
in details with N-body simulations by \cite{Hilz+13}. Here
three fiducial models with mass-ratio of 1:1, 5:1 and 10:1 are
analysed to find that, at different final stellar masses, the
equal-mass mergers produce a smaller size increase of multiple
minor mergers. In particular, the variation of \Re\ with respect
to the initial radius, $\Re / R_{\rm 0}$, in terms of the
variation of \mst\ with respect to the initial stellar mass, $\mst
/ M_{\rm 0}$, is found to be $\Re/R_{\rm 0} \propto$ $(\mst/M_{\rm
0})^{0.91}$ for the equal-mass merger and $\propto (\mst/M_{\rm
0})^{2.4}$ for the minor mergers. These latter predictions on the
\Re\ and \mst\ accretion have been found to be consistent with
observations (\citealt{vanDokkum+10}).

Taking these results into account we have constructed some
toy-models assuming that $M_{\rm DM} \propto \Mvir r^{\eta}$
around \Re, with $\eta \sim 2$ for a standard NFW and $\eta \sim
1.2$ for a contracted NFW, hereafter AC+NFW (according with
\citealt{Boylan-Kolchin+05}). Consistently with \cite{Hilz+13}, we
have also taken the average evolution of \Re\ in terms of \mst\
evolution for the equal-mass merging (i.e. $\Re/R_{\rm 0} \propto
(\mst/M_{\rm 0})^{0.91}$) and minor merging (i.e. $\Re/R_{\rm 0}
\propto (\mst/M_{\rm 0})^{2.4}$) with the further assumption that
the variation of the virial mass almost follow the one of the
stellar mass, i.e. $\delta \Mvir \approx \delta \mst$. This
intrinsically reflects the hypothesis that the systems concurring
to mergings all have some constant $\Mvir / \mst$, which is
reasonable for most of the stellar mass range covered by our
sample.

These simplified models have the advantage of dealing with a
limited number of parameters, but still providing a quantitative
assessment of the impact of the evolution scenario on the observed
DM fractions. In particular, we need to clarify whether the
observed DM fraction evolution is compatible with a variation with
$z$ of \Re\ and \TtoSM, or rather it is the IMF variation with $z$
the only viable explanation of the observed correlations.

In Fig. \ref{fig:evolution} we show the predicted evolved tracks
for the the median \Re\ and \TtoSM\ of the EDisCS dataset
corresponding to a median $\log\mst/\Msun = 11$. We have
considered the evolution tracks related to different merging
types, according to the list in Table \ref{tab:tab3}. The major
merging tracks are shown as blue lines with dots/squares
indicating the events corresponding to $x\times M_{\rm 0}$
($x=0,1,2, . . .,n$) mass increments, while the minor merging
tracks are shown by yellow lines where dots/squares indicate the
$x\times M_{\rm 0}$ with $x=0,0.5,1,. . .,n$. Dots and squares are
for NFW and AC+NFW profiles, respectively.

Comparing the model predictions with local results, we see that
after a single major merger is not possible to reproduce the local
correlations (blue lines in \Re--\mst\ and \TtoSM--\mst). Adopting
a NFW profile ({\tt Major NFW}), after $\sim 3$ major mergers the
high$-z$ galaxy moves along the track to overlap to the local
observed \TtoSM, while it would need more merging events if a
AC+NFW is used ({\tt Major AC+NFW}, open squares). In both cases,
though, the toy-model predictions do not match the average local
\Re--\mst\ relations, showing a deficiency of size growth expected
for major merging events.

On the contrary, after a few minor mergers (yellow lines) which accrete $\sim
0.5-1.5$ of the initial \mst\ (depending on the DM profile adopted), both
\Re--\mst\ and \TtoSM--\mst\ are matched (see {\tt Minor NFW}
and {\tt Minor AC+NFW}).

To summarise the behaviour of our toy-models: a) the impact of
minor merging on \Re\ and \TtoSM\ is stronger than the one of
major merging at fixed accreted \mst, and the latter do not seem
to account for the observed \Re\ evolution; b) uncontracted NFW
models provide larger \TtoSM .

\begin{table}
\centering \caption{Typical parameters of toy-models for major and
minor mergings. See text for details. The predictions for \Re\ and
\TtoSM\ are plotted in Fig. \ref{fig:evolution}: { {\tt
Major NFW} (blue lines and dots), {\tt Minor NFW} (yellow lines
and dots), {\tt Major AC+NFW} (blue lines and squares), {\tt Minor
AC+NFW} (yellow lines and squares). See text and caption of Fig.
\ref{fig:evolution} for further details.} }\label{tab:tab3}
\begin{tabular}{lcccc} 
\hline
\rm Model & $\mst/M_{\rm 0}$ & Merging type & $\eta$ \\ 
\hline
{\tt Major NFW} & 1,2,3 & Major & 2 \\ 
{\tt Minor NFW} & 0.5,1,1.5,... & Minor & 2 \\ 

{\tt Major AC+NFW} & 1,2,3 & Major & 1.2 \\ 
{\tt Minor AC+NFW} & 0.5,1,1.5,... & Minor & 1.2 \\ 
\hline
\end{tabular}
\end{table}

For simplicity, we have not analyzed here the case of a \Mvir\
which changes less (more) than \mst\, i.e. $\delta \Mvir < \delta
\mst$ ($\delta \Mvir > \delta \mst$). For example, if $\delta
\Mvir > \delta \mst$ then \Mdyn\ and \TtoSM\ get larger.

In Sect. \ref{sec:DM_corr} we have shown that \TtoSM\ are slightly
smaller in field than in cluster galaxies, but this difference is
not statistical significant. For galaxies living within dense
environments, the chance to interact with neighboring galaxies is
very high, but for galaxies in the field merging is a less common
event. Thus, the only ways to put EDisCS field galaxies and local
objects within a coherent scenario are: a) they will fall in the
following 7 Gyrs within a cluster or a group of galaxies, and
merge with other galaxies, b) they are atypical objects which set
out of the median local relations, c) any other phenomenon is
working to puffing up the galaxy size (as AGN feedback;
\citealt{Fan+08, Fan+10}) and consequently the \TtoSM.

\section{Conclusions}\label{sec:conclusions}

We have analyzed the central DM content in a sample of high-$z$
($\sim 0.4-0.8$)  ETGs from the EDisCS survey, parameterized in
this paper through the  total-to-stellar mass ratio, \TtoSM,
(calculated at $r = 1\, \rm \Re$), and compared the results with
local galaxy samples (SPIDER, \citealt{SPIDER-VI}; \atlas3d,
\citealt{Cappellari+11_ATLAS3D_I}; \citealt{Tortora+09}).

{We have shown that local correlations between \TtoSM\ and \Re,
\mst\ and \sige\ are fairly independent of the sample adopted and
almost all conserved at high$-z$ (see Fig. \ref{fig:fdm}). In
fact, high$-z$ galaxies are more DM dominated at larger \Re\ and
\sige\ (i.e. they follow a positive correlation). For this sample,
this is a clear indication that the main driver in the observed
positive correlations is the effective radius, consistently with
local galaxies as found in \cite{NRT10} (see also
\citealt{Auger+10_SLACSX}; \citealt{Tortora+10lensing};
\citealt{SPIDER-VI}). A less significant negative correlation is
found in terms of stellar mass, in the opposite direction to the
correlations of the local galaxies.

As seen in Figs. \ref{fig:size-mass-fdm} and \ref{fig:fdm}, an
overall offset of the \TtoSM\ from high$-$ to low$-z$ galaxies is
evident, suggesting an evolution of the DM fraction within \Re.}
In particular, we find that at fixed \Re\, the EDisCS sample
presents \TtoSM\ of $\sim 0.15-0.25$ dex smaller than local
galaxies. The change in \TtoSM\ is stronger at \mst\ and \sige\
fixed, being $\sim 0.3-0.35$ dex and $\sim 0.5-0.6$ dex,
respectively. At fixed \mst\ a clear variation of the \TtoSM\ as a
function of \Re\ is found from high$-$ to low$-z$ galaxies (Fig.
\ref{fig:size-mass-fdm}), but not in \sige\ (Fig.
\ref{fig:size-sigma-mass_2}). The overall evolution of the DM
fraction with redshift (see Fig. \ref{fig:final}) is consistent
with the results found in \cite{Faure+11}, but in contrast with
the opposite trend found by \cite{Ruff+11} (within $\Re/2$), which
find larger DM fractions at high-$z$, and inconsistent with the
absence of evolution in \cite{Bezanson+13}. Our findings are
consistent with results in \cite{Beifiori+14} which, using
SDSS-III/BOSS and SDSS-II datasets, find smaller DM fractions in
higher-z galaxies, but the trend with redshift is shallower than
ours. Instead, we do not agree with the absence of evolution found
in \cite{BNE14} for a sample of quiescent galaxies at $z
> 1$.

We find that, on average, the inferred \fdm\ from a Salpeter IMF
(or bottom-heavier IMFs in general) are disfavored with respect to
a Chabrier IMF, mainly at high redshift and in low-\sige\ systems.
Thus, we can reproduce our \fdm\ assuming a Chabrier/Salpeter IMF
at high/low redshift, which would point to an IMF evolution with
redshift. This result seems to be in contrast with the recent
findings in \cite{Shetty_Cappellari14}, which find a Salpeter IMF
in $z\sim 0.7-0.8$ massive galaxies. However, their models assume
a constant-\ML\ profile, and no-DM content in the central regions
(i.e. the mass is only in stars) which might have biased the
overall stellar \ML\ as their dynamical \ML\ and the inferred
\Yst\ are just upper limits (\citealt{TRN13_SPIDER_IMF}).
Furthermore their sample is composed by very massive ($\mst \gsim
10^{11}\, \rm \Msun$) galaxies with very high-$\sigma$ (most of
the systems have $\sigma > 200 \, \rm km/s$). Taking all these
specifics of their sample, their results are substantially
consistent with ours, since a Salpeter IMF is allowed by the
EDisCS systems at similar velocity dispersions, providing nearly
null DM fractions consistent with their assumption (see bottom
panels in Fig. \ref{fig:fdm})

We have briefly discussed the possibility that the IMF is
non-universal, not only as a function of redshift, but that it can
also vary with the mass/central velocity dispersion (see, e.g.,
\citealt{TRN13_SPIDER_IMF}). The impact on our results of a more
realistic IMF variation with \sige\ from local observations
(\citealt{TRN13_SPIDER_IMF}) has been analyzed, and assumed to be
valid and the same at all redshifts. This has been shown to
produce a smaller number of negative DM fraction with respect to a
Salpeter IMF. However, a combined evolution with mass/\sige\ and
redshift would completely solve the issue of negative \fdm .
Although most of stars are already in place at $z<1$, a change of
IMF across the time could be produced by two different processes:
a) new stars formed in the core during a wet merging process,
which also produce positive age gradients (e.g.,
\citealt{Hopkins+08_DELGN_I}; \citealt{Tortora+10CG}) and
``higher-mass'' IMF (\citealt{NRT10}) in young and massive local
ETGs, or b) by stars from both merging galaxies characterized by
two different IMFs, which can combine to modify the cumulative IMF
of the remnant. Unfortunately, the net effect of these processes
on the final IMF is not yet clear and in most cases the
combination of a ``higher-'' and a ``lower-mass'' IMF would
produce a final diluted IMF. Therefore, it is also plausible to
scan alternative mass modelling and mechanisms to explain the
negative \fdm\ at high--$z$ and in low-\sige\ ETGs, as i) adopting
realistic and viable DM halo profiles, ii) the effect of some
dissipative processes that might alter the standard DM
distribution like the adiabatic contraction (\citealt{NRT10};
\citealt{TRN13_SPIDER_IMF}) or iii) some other DM flavor (e.g.,
some warm DM, \citealt{Schneider+12}).

However, a detailed analysis of the IMF and DM halo evolution
across the time is beyond the scopes of this paper and will be
addressed in future, assuming more complex mass modelling
(\citealt{TRN13_SPIDER_IMF}; \citealt{Tortora+14_MOND}). From the
theoretic point of view, this evolution would pose a further open
question: which circumstances might have caused the IMF variation
across time at any given \mst ?. This is a new fundamental
question which has to be added to the IMF evidences of
non-universality (\citealt{Conroy_vanDokkum12b};
\citealt{Cappellari+12}; \citealt{Spiniello+12};
\citealt{Dutton+13}; \citealt{Ferreras+13};
\citealt{Goudfrooij_Kruijssen13};
\citealt{LaBarbera+13_SPIDERVIII_IMF}; \citealt{TRN13_SPIDER_IMF};
\citealt{Weidner+13_giant_ell}; \citealt{Tortora+14_MOND}).

We have finally investigated our results with a fixed IMF within
galaxy formation scenarios and found that simple passive evolution
is not able to reproduce local results. On the contrary, the
galaxy merging scenario allow to reproduce the growth of the \Re\
and the \TtoSM\ with $z$. We have used toy-models which take into
account size and mass accretion from minor- and major-mergings to
show that a single major merger is not able to reproduce local
correlations, while many minor mergers work better (e.g.,
\citealt{Hilz+13}; \citealt{BNE14}).

Of course, in order to have a firmer assessment on the actual DM
content evolution with redshift, much larger samples are needed,
with full spectroscopic and photometric information. In
particular, gravitational lensing can provide us very robust mass
estimates (e.g., \citealt{Barnabe+09}; \citealt{Covone+09};
\citealt{Auger+10_SLACSX}; \citealt{Tortora+10lensing};
\citealt{Barnabe+11}; \citealt{Spiniello+11}). Thus, survey
projects like SLACS or COSMOS (e.g. \citealt{Faure+08_COSMOS}) or
the ongoing ESO public surveys with VST telescope (such as KiDS)
together with spectroscopic surveys (e.g., BOSS/SDSS,
\citealt{Thomas+13_BOSS, Beifiori+14} and GAMA,
\citealt{Baldry+10_GAMA}) are fundamental to collect a large
sample of galaxies, that span a wide range of luminosity and
redshifts, in order to probe more massive galaxies in the universe
and their evolutive history, by means of the description of the
different history of stellar and DM and their interplay in the
inner and outer regions of galaxies.


\section*{Acknowledgments}

We thank the referee for his/her comments which helped to improve
the manuscript. CT has received funding from the European Union
Seventh Framework Programme (FP7/2007-2013) under grant agreement
n. 267251.


\bibliographystyle{mn2e}   



\begin{thebibliography}{120}
\expandafter\ifx\csname
natexlab\endcsname\relax\def\natexlab#1{#1}\fi

\bibitem[{{Abazajian} {et~al}\mbox{.}(2003){Abazajian}, {Adelman-McCarthy},
  {Ag{\"u}eros}, {Allam}, {Anderson}, {Annis}, {Bahcall}, {Baldry}, {Bastian},
  {Berlind}, {Bernardi}, {Blanton}, {Blythe}, {Bochanski}, {Boroski},
  {Brewington}, {Briggs}, {Brinkmann}, {Brunner}, {Budav{\'a}ri}, {Carey},
  {Carr}, {Castander}, {Chiu}, {Collinge}, {Connolly}, {Covey}, {Csabai},
  {Dalcanton}, {Dodelson}, {Doi}, {Dong}, {Eisenstein}, {Evans}, {Fan},
  {Feldman}, {Finkbeiner}, {Friedman}, {Frieman}, {Fukugita}, {Gal},
  {Gillespie}, {Glazebrook}, {Gonzalez}, {Gray}, {Grebel}, {Grodnicki}, {Gunn},
  {Gurbani}, {Hall}, {Hao}, {Harbeck}, {Harris}, {Harris}, {Harvanek},
  {Hawley}, {Heckman}, {Helmboldt}, {Hendry}, {Hennessy}, {Hindsley}, {Hogg},
  {Holmgren}, {Holtzman}, {Homer}, {Hui}, {Ichikawa}, {Ichikawa}, {Inkmann},
  {Ivezi{\'c}}, {Jester}, {Johnston}, {Jordan}, {Jordan}, {Jorgensen},
  {Juri{\'c}}, {Kauffmann}, {Kent}, {Kleinman}, {Knapp}, {Kniazev}, {Kron},
  {Krzesi{\'n}ski}, {Kunszt}, {Kuropatkin}, {Lamb}, {Lampeitl}, {Laubscher},
  {Lee}, {Leger}, {Li}, {Lidz}, {Lin}, {Loh}, {Long}, {Loveday}, {Lupton},
  {Malik}, {Margon}, {McGehee}, {McKay}, {Meiksin}, {Miknaitis}, {Moorthy},
  {Munn}, {Murphy}, {Nakajima}, {Narayanan}, {Nash}, {Neilsen}, {Newberg},
  {Newman}, {Nichol}, {Nicinski}, {Nieto-Santisteban}, {Nitta}, {Odenkirchen},
  {Okamura}, {Ostriker}, {Owen}, {Padmanabhan}, {Peoples}, {Pier}, {Pindor},
  {Pope}, {Quinn}, {Rafikov}, {Raymond}, {Richards}, {Richmond}, {Rix},
  {Rockosi}, {Schaye}, {Schlegel}, {Schneider}, {Schroeder}, {Scranton},
  {Sekiguchi}, {Seljak}, {Sergey}, {Sesar}, {Sheldon}, {Shimasaku}, {Siegmund},
  {Silvestri}, {Sinisgalli}, {Sirko}, {Smith}, {Smol{\v c}i{\'c}}, {Snedden},
  {Stebbins}, {Steinhardt}, {Stinson}, {Stoughton}, {Strateva}, {Strauss},
  {SubbaRao}, {Szalay}, {Szapudi}, {Szkody}, {Tasca}, {Tegmark}, {Thakar},
  {Tremonti}, {Tucker}, {Uomoto}, {Vanden Berk}, {Vandenberg}, {Vogeley},
  {Voges}, {Vogt}, {Walkowicz}, {Weinberg}, {West}, {White}, {Wilhite},
  {Willman}, {Xu}, {Yanny}, {Yarger}, {Yasuda}, {Yip}, {Yocum}, {York},
  {Zakamska}, {Zehavi}, {Zheng}, {Zibetti}, \& {Zucker}}]{SDSS_DR1}
{Abazajian} K. {et~al.}, 2003, \aj, 126, 2081

\bibitem[{{Abazajian} {et~al}\mbox{.}(2009){Abazajian}, {Adelman-McCarthy},
  {Ag{\"u}eros}, {Allam}, {Allende Prieto}, {An}, {Anderson}, {Anderson},
  {Annis}, {Bahcall}, \& et~al.}]{SDSS_DR7}
{Abazajian} K.~N. {et~al.}, 2009, \apjs, 182, 543

\bibitem[{{Adelman-McCarthy} {et~al}\mbox{.}(2008){Adelman-McCarthy},
  {Ag{\"u}eros}, {Allam}, {Allende Prieto}, {Anderson}, {Anderson}, {Annis},
  {Bahcall}, {Bailer-Jones}, {Baldry}, {Barentine}, {Bassett}, {Becker},
  {Beers}, {Bell}, {Berlind}, {Bernardi}, {Blanton}, {Bochanski}, {Boroski},
  {Brinchmann}, {Brinkmann}, {Brunner}, {Budav{\'a}ri}, {Carliles}, {Carr},
  {Castander}, {Cinabro}, {Cool}, {Covey}, {Csabai}, {Cunha}, {Davenport},
  {Dilday}, {Doi}, {Eisenstein}, {Evans}, {Fan}, {Finkbeiner}, {Friedman},
  {Frieman}, {Fukugita}, {G{\"a}nsicke}, {Gates}, {Gillespie}, {Glazebrook},
  {Gray}, {Grebel}, {Gunn}, {Gurbani}, {Hall}, {Harding}, {Harvanek}, {Hawley},
  {Hayes}, {Heckman}, {Hendry}, {Hindsley}, {Hirata}, {Hogan}, {Hogg}, {Hyde},
  {Ichikawa}, {Ivezi{\'c}}, {Jester}, {Johnson}, {Jorgensen}, {Juri{\'c}},
  {Kent}, {Kessler}, {Kleinman}, {Knapp}, {Kron}, {Krzesinski}, {Kuropatkin},
  {Lamb}, {Lampeitl}, {Lebedeva}, {Lee}, {Leger}, {L{\'e}pine}, {Lima}, {Lin},
  {Long}, {Loomis}, {Loveday}, {Lupton}, {Malanushenko}, {Malanushenko},
  {Mandelbaum}, {Margon}, {Marriner}, {Mart{\'{\i}}nez-Delgado}, {Matsubara},
  {McGehee}, {McKay}, {Meiksin}, {Morrison}, {Munn}, {Nakajima}, {Neilsen},
  {Newberg}, {Nichol}, {Nicinski}, {Nieto-Santisteban}, {Nitta}, {Okamura},
  {Owen}, {Oyaizu}, {Padmanabhan}, {Pan}, {Park}, {Peoples}, {Pier}, {Pope},
  {Purger}, {Raddick}, {Re Fiorentin}, {Richards}, {Richmond}, {Riess}, {Rix},
  {Rockosi}, {Sako}, {Schlegel}, {Schneider}, {Schreiber}, {Schwope}, {Seljak},
  {Sesar}, {Sheldon}, {Shimasaku}, {Sivarani}, {Smith}, {Snedden}, {Steinmetz},
  {Strauss}, {SubbaRao}, {Suto}, {Szalay}, {Szapudi}, {Szkody}, {Tegmark},
  {Thakar}, {Tremonti}, {Tucker}, {Uomoto}, {Vanden Berk}, {Vandenberg},
  {Vidrih}, {Vogeley}, {Voges}, {Vogt}, {Wadadekar}, {Weinberg}, {West},
  {White}, {Wilhite}, {Yanny}, {Yocum}, {York}, {Zehavi}, \&
  {Zucker}}]{SDSS_DR6}
{Adelman-McCarthy} J.~K. {et~al.}, 2008, \apjs, 175, 297

\bibitem[{{Auger} {et~al}\mbox{.}(2009){Auger}, {Treu}, {Bolton}, {Gavazzi},
  {Koopmans}, {Marshall}, {Bundy}, \& {Moustakas}}]{Auger+09_SLACSIX}
{Auger} M.~W., {Treu} T., {Bolton} A.~S., {Gavazzi} R., {Koopmans}
L.~V.~E.,
  {Marshall} P.~J., {Bundy} K., {Moustakas} L.~A., 2009, \apj, 705, 1099

\bibitem[{{Auger} {et~al}\mbox{.}(2010){Auger}, {Treu}, {Bolton}, {Gavazzi},
  {Koopmans}, {Marshall}, {Moustakas}, \& {Burles}}]{Auger+10_SLACSX}
{Auger} M.~W., {Treu} T., {Bolton} A.~S., {Gavazzi} R., {Koopmans}
L.~V.~E.,
  {Marshall} P.~J., {Moustakas} L.~A., {Burles} S., 2010, \apj, 724, 511

\bibitem[{{Baldry} {et~al}\mbox{.}(2010){Baldry}, {Robotham}, {Hill}, {Driver},
  {Liske}, {Norberg}, {Bamford}, {Hopkins}, {Loveday}, {Peacock}, {Cameron},
  {Croom}, {Cross}, {Doyle}, {Dye}, {Frenk}, {Jones}, {van Kampen}, {Kelvin},
  {Nichol}, {Parkinson}, {Popescu}, {Prescott}, {Sharp}, {Sutherland},
  {Thomas}, \& {Tuffs}}]{Baldry+10_GAMA}
{Baldry} I.~K. {et~al.}, 2010, \mnras, 404, 86

\bibitem[{{Barnab{\`e}} {et~al}\mbox{.}(2011){Barnab{\`e}}, {Czoske},
  {Koopmans}, {Treu}, \& {Bolton}}]{Barnabe+11}
{Barnab{\`e}} M., {Czoske} O., {Koopmans} L.~V.~E., {Treu} T.,
{Bolton} A.~S.,
  2011, \mnras, 415, 2215

\bibitem[{{Barnab{\`e}} {et~al}\mbox{.}(2009){Barnab{\`e}}, {Czoske},
  {Koopmans}, {Treu}, {Bolton}, \& {Gavazzi}}]{Barnabe+09}
{Barnab{\`e}} M., {Czoske} O., {Koopmans} L.~V.~E., {Treu} T.,
{Bolton} A.~S.,
  {Gavazzi} R., 2009, \mnras, 399, 21

\bibitem[{{Beifiori} {et~al}\mbox{.}(2014){Beifiori}, {Thomas}, {Maraston},
  {Steele}, {Masters}, {Pforr}, {Saglia}, {Bender}, {Tojeiro}, {Chen},
  {Bolton}, {Brownstein}, {Johansson}, {Leauthaud}, {Nichol}, {Schneider},
  {Senger}, {Skibba}, {Wake}, {Pan}, {Snedden}, {Bizyaev}, {Brewington},
  {Malanushenko}, {Malanushenko}, {Oravetz}, {Simmons}, {Shelden}, \&
  {Ebelke}}]{Beifiori+14}
{Beifiori} A. {et~al.}, 2014, ArXiv e-prints

\bibitem[{{Bell} \& {de Jong}(2001)}]{Bell_deJong01}
{Bell} E.~F., {de Jong} R.~S., 2001, \apj, 550, 212

\bibitem[{{Belli} {et~al}\mbox{.}(2014){Belli}, {Newman}, \& {Ellis}}]{BNE14}
{Belli} S., {Newman} A.~B., {Ellis} R.~S., 2014, \apj, 783, 117

\bibitem[{{Benson} {et~al}\mbox{.}(2000){Benson}, {Cole}, {Frenk}, {Baugh}, \&
  {Lacey}}]{Benson+00}
{Benson} A.~J., {Cole} S., {Frenk} C.~S., {Baugh} C.~M., {Lacey}
C.~G., 2000,
  \mnras, 311, 793

\bibitem[{{Bezanson} {et~al}\mbox{.}(2013){Bezanson}, {van Dokkum}, {van de
  Sande}, {Franx}, {Leja}, \& {Kriek}}]{Bezanson+13}
{Bezanson} R., {van Dokkum} P.~G., {van de Sande} J., {Franx} M.,
{Leja} J.,
  {Kriek} M., 2013, \apjl, 779, L21

\bibitem[{{Bolton} {et~al}\mbox{.}(2008){Bolton}, {Burles}, {Koopmans}, {Treu},
  {Gavazzi}, {Moustakas}, {Wayth}, \& {Schlegel}}]{Bolton+08_SLACSV}
{Bolton} A.~S., {Burles} S., {Koopmans} L.~V.~E., {Treu} T.,
{Gavazzi} R.,
  {Moustakas} L.~A., {Wayth} R., {Schlegel} D.~J., 2008, \apj, 682, 964

\bibitem[{{Bolton} {et~al}\mbox{.}(2006){Bolton}, {Burles}, {Koopmans}, {Treu},
  \& {Moustakas}}]{Bolton+06_SLACSI}
{Bolton} A.~S., {Burles} S., {Koopmans} L.~V.~E., {Treu} T.,
{Moustakas} L.~A.,
  2006, \apj, 638, 703

\bibitem[{{Boylan-Kolchin} {et~al}\mbox{.}(2005){Boylan-Kolchin}, {Ma}, \&
  {Quataert}}]{Boylan-Kolchin+05}
{Boylan-Kolchin} M., {Ma} C.-P., {Quataert} E., 2005, \mnras, 362,
184

\bibitem[{{Bruzual} \& {Charlot}(2003)}]{BC03}
{Bruzual} G., {Charlot} S., 2003, \mnras, 344, 1000

\bibitem[{{Bullock} {et~al}\mbox{.}(2001){Bullock}, {Kolatt}, {Sigad},
  {Somerville}, {Kravtsov}, {Klypin}, {Primack}, \& {Dekel}}]{Bullock+01}
{Bullock} J.~S., {Kolatt} T.~S., {Sigad} Y., {Somerville} R.~S.,
{Kravtsov}
  A.~V., {Klypin} A.~A., {Primack} J.~R., {Dekel} A., 2001, \mnras, 321, 559

\bibitem[{{Cappellari} {et~al}\mbox{.}(2006){Cappellari}, {Bacon}, {Bureau},
  {Damen}, {Davies}, {de Zeeuw}, {Emsellem}, {Falc{\'o}n-Barroso},
  {Krajnovi{\'c}}, {Kuntschner}, {McDermid}, {Peletier}, {Sarzi}, {van den
  Bosch}, \& {van de Ven}}]{Cappellari+06}
{Cappellari} M. {et~al.}, 2006, \mnras, 366, 1126

\bibitem[{{Cappellari} {et~al}\mbox{.}(2011){Cappellari}, {Emsellem},
  {Krajnovi{\'c}}, {McDermid}, {Scott}, {Verdoes Kleijn}, {Young}, {Alatalo},
  {Bacon}, {Blitz}, {Bois}, {Bournaud}, {Bureau}, {Davies}, {Davis}, {de
  Zeeuw}, {Duc}, {Khochfar}, {Kuntschner}, {Lablanche}, {Morganti}, {Naab},
  {Oosterloo}, {Sarzi}, {Serra}, \& {Weijmans}}]{Cappellari+11_ATLAS3D_I}
{Cappellari} M. {et~al.}, 2011, \mnras, 413, 813

\bibitem[{{Cappellari} {et~al}\mbox{.}(2012){Cappellari}, {McDermid},
  {Alatalo}, {Blitz}, {Bois}, {Bournaud}, {Bureau}, {Crocker}, {Davies},
  {Davis}, {de Zeeuw}, {Duc}, {Emsellem}, {Khochfar}, {Krajnovi{\'c}},
  {Kuntschner}, {Lablanche}, {Morganti}, {Naab}, {Oosterloo}, {Sarzi}, {Scott},
  {Serra}, {Weijmans}, \& {Young}}]{Cappellari+12}
{Cappellari} M. {et~al.}, 2012, \nat, 484, 485

\bibitem[{{Cappellari} {et~al}\mbox{.}(2013{\natexlab{a}}){Cappellari},
  {McDermid}, {Alatalo}, {Blitz}, {Bois}, {Bournaud}, {Bureau}, {Crocker},
  {Davies}, {Davis}, {de Zeeuw}, {Duc}, {Emsellem}, {Khochfar},
  {Krajnovi{\'c}}, {Kuntschner}, {Morganti}, {Naab}, {Oosterloo}, {Sarzi},
  {Scott}, {Serra}, {Weijmans}, \& {Young}}]{Cappellari+13_ATLAS3D_XX}
{Cappellari} M. {et~al.}, 2013{\natexlab{a}}, \mnras, 432, 1862

\bibitem[{{Cappellari} {et~al}\mbox{.}(2013{\natexlab{b}}){Cappellari},
  {Scott}, {Alatalo}, {Blitz}, {Bois}, {Bournaud}, {Bureau}, {Crocker},
  {Davies}, {Davis}, {de Zeeuw}, {Duc}, {Emsellem}, {Khochfar},
  {Krajnovi{\'c}}, {Kuntschner}, {McDermid}, {Morganti}, {Naab}, {Oosterloo},
  {Sarzi}, {Serra}, {Weijmans}, \& {Young}}]{Cappellari+13_ATLAS3D_XV}
{Cappellari} M. {et~al.}, 2013{\natexlab{b}}, \mnras, 432, 1709

\bibitem[{{Cardone} {et~al}\mbox{.}(2011){Cardone}, {Del Popolo}, {Tortora}, \&
  {Napolitano}}]{Cardone+11SIM}
{Cardone} V.~F., {Del Popolo} A., {Tortora} C., {Napolitano}
N.~R., 2011,
  \mnras, 416, 1822

\bibitem[{{Cardone} \& {Tortora}(2010)}]{CT10}
{Cardone} V.~F., {Tortora} C., 2010, \mnras, 409, 1570

\bibitem[{{Cardone} {et~al}\mbox{.}(2009){Cardone}, {Tortora}, {Molinaro}, \&
  {Salzano}}]{Cardone+09}
{Cardone} V.~F., {Tortora} C., {Molinaro} R., {Salzano} V., 2009,
\aap, 504,
  769

\bibitem[{{Cenarro} \& {Trujillo}(2009)}]{Cenarro_Trujillo09}
{Cenarro} A.~J., {Trujillo} I., 2009, \apjl, 696, L43

\bibitem[{{Chabrier}(2001)}]{Chabrier01}
{Chabrier} G., 2001, \apj, 554, 1274

\bibitem[{{Chae} {et~al}\mbox{.}(2014){Chae}, {Bernardi}, \&
  {Kravtsov}}]{Chae+14}
{Chae} K.-H., {Bernardi} M., {Kravtsov} A.~V., 2014, \mnras, 437,
3670

\bibitem[{{Cid Fernandes} {et~al}\mbox{.}(2005){Cid Fernandes}, {Mateus},
  {Sodr{\'e}}, {Stasi{\'n}ska}, \& {Gomes}}]{CidFernandes+05}
{Cid Fernandes} R., {Mateus} A., {Sodr{\'e}} L., {Stasi{\'n}ska}
G., {Gomes}
  J.~M., 2005, \mnras, 358, 363

\bibitem[{{Conroy} \& {van Dokkum}(2012)}]{Conroy_vanDokkum12b}
{Conroy} C., {van Dokkum} P.~G., 2012, \apj, 760, 71

\bibitem[{{Conroy} \& {Wechsler}(2009)}]{CW09}
{Conroy} C., {Wechsler} R.~H., 2009, \apj, 696, 620

\bibitem[{{Covone} {et~al}\mbox{.}(2009){Covone}, {Paolillo}, {Napolitano},
  {Capaccioli}, {Longo}, {Kneib}, {Jullo}, {Richard}, {Khovanskaya}, {Sazhin},
  {Grogin}, \& {Schreier}}]{Covone+09}
{Covone} G. {et~al.}, 2009, \apj, 691, 531

\bibitem[{{de Vaucouleurs}(1948)}]{deVauc48}
{de Vaucouleurs} G., 1948, Annales d'Astrophysique, 11, 247

\bibitem[{{Dutton} {et~al}\mbox{.}(2013){Dutton}, {Macci{\`o}}, {Mendel}, \&
  {Simard}}]{Dutton+13}
{Dutton} A.~A., {Macci{\`o}} A.~V., {Mendel} J.~T., {Simard} L.,
2013, \mnras,
  432, 2496

\bibitem[{{Dutton} \& {Treu}(2014)}]{Dutton_Treu14}
{Dutton} A.~A., {Treu} T., 2014, \mnras, 438, 3594

\bibitem[{{Fan} {et~al}\mbox{.}(2010){Fan}, {Lapi}, {Bressan}, {Bernardi}, {De
  Zotti}, \& {Danese}}]{Fan+10}
{Fan} L., {Lapi} A., {Bressan} A., {Bernardi} M., {De Zotti} G.,
{Danese} L.,
  2010, \apj, 718, 1460

\bibitem[{{Fan} {et~al}\mbox{.}(2008){Fan}, {Lapi}, {De Zotti}, \&
  {Danese}}]{Fan+08}
{Fan} L., {Lapi} A., {De Zotti} G., {Danese} L., 2008, \apjl, 689,
L101

\bibitem[{{Faure} {et~al}\mbox{.}(2011){Faure}, {Anguita}, {Alloin}, {Bundy},
  {Finoguenov}, {Leauthaud}, {Knobel}, {Kneib}, {Jullo}, {Ilbert}, {Koekemoer},
  {Capak}, {Scoville}, \& {Tasca}}]{Faure+11}
{Faure} C. {et~al.}, 2011, \aap, 529, A72

\bibitem[{{Faure} {et~al}\mbox{.}(2008){Faure}, {Kneib}, {Covone}, {Tasca},
  {Leauthaud}, {Capak}, {Jahnke}, {Smolcic}, {de la Torre}, {Ellis},
  {Finoguenov}, {Koekemoer}, {Le Fevre}, {Massey}, {Mellier}, {Refregier},
  {Rhodes}, {Scoville}, {Schinnerer}, {Taylor}, {Van Waerbeke}, \&
  {Walcher}}]{Faure+08_COSMOS}
{Faure} C. {et~al.}, 2008, \apjs, 176, 19

\bibitem[{{Ferreras} {et~al}\mbox{.}(2013){Ferreras}, {La Barbera}, {de la
  Rosa}, {Vazdekis}, {de Carvalho}, {Falc{\'o}n-Barroso}, \&
  {Ricciardelli}}]{Ferreras+13}
{Ferreras} I., {La Barbera} F., {de la Rosa} I.~G., {Vazdekis} A.,
{de
  Carvalho} R.~R., {Falc{\'o}n-Barroso} J., {Ricciardelli} E., 2013, \mnras,
  429, L15

\bibitem[{{Gavazzi} {et~al}\mbox{.}(2007){Gavazzi}, {Treu}, {Rhodes},
  {Koopmans}, {Bolton}, {Burles}, {Massey}, \&
  {Moustakas}}]{Gavazzi+07_SLACSIV}
{Gavazzi} R., {Treu} T., {Rhodes} J.~D., {Koopmans} L.~V.~E.,
{Bolton} A.~S.,
  {Burles} S., {Massey} R.~J., {Moustakas} L.~A., 2007, \apj, 667, 176

\bibitem[{{Gerhard} {et~al}\mbox{.}(2001){Gerhard}, {Kronawitter}, {Saglia}, \&
  {Bender}}]{Gerhard+01}
{Gerhard} O., {Kronawitter} A., {Saglia} R.~P., {Bender} R., 2001,
\aj, 121,
  1936

\bibitem[{{Gnedin} {et~al}\mbox{.}(2004){Gnedin}, {Kravtsov}, {Klypin}, \&
  {Nagai}}]{Gnedin+04}
{Gnedin} O.~Y., {Kravtsov} A.~V., {Klypin} A.~A., {Nagai} D.,
2004, \apj, 616,
  16

\bibitem[{{Gnedin} {et~al}\mbox{.}(2007){Gnedin}, {Weinberg}, {Pizagno},
  {Prada}, \& {Rix}}]{Gnedin+07}
{Gnedin} O.~Y., {Weinberg} D.~H., {Pizagno} J., {Prada} F., {Rix}
H.-W., 2007,
  \apj, 671, 1115

\bibitem[{{Goudfrooij} \& {Kruijssen}(2013)}]{Goudfrooij_Kruijssen13}
{Goudfrooij} P., {Kruijssen} J.~M.~D., 2013, \apj, 762, 107

\bibitem[{{Goudfrooij} \& {Kruijssen}(2014)}]{Goudfrooij_Kruijssen14}
{Goudfrooij} P., {Kruijssen} J.~M.~D., 2014, \apj, 780, 43

\bibitem[{{Graves} {et~al}\mbox{.}(2009){Graves}, {Faber}, \&
  {Schiavon}}]{Graves+09}
{Graves} G.~J., {Faber} S.~M., {Schiavon} R.~P., 2009, \apj, 698,
1590

\bibitem[{{Grillo}(2010)}]{Grillo10}
{Grillo} C., 2010, \apj, 722, 779

\bibitem[{{Grillo} \& {Gobat}(2010)}]{Grillo_Cobat10}
{Grillo} C., {Gobat} R., 2010, \mnras, 402, L67

\bibitem[{{Grillo} {et~al}\mbox{.}(2009){Grillo}, {Gobat}, {Lombardi}, \&
  {Rosati}}]{Grillo+09}
{Grillo} C., {Gobat} R., {Lombardi} M., {Rosati} P., 2009, \aap,
501, 461

\bibitem[{{Heymans} {et~al}\mbox{.}(2006){Heymans}, {Bell}, {Rix}, {Barden},
  {Borch}, {Caldwell}, {McIntosh}, {Meisenheimer}, {Peng}, {Wolf}, {Beckwith},
  {H{\"a}u{\ss}ler}, {Jahnke}, {Jogee}, {S{\'a}nchez}, {Somerville}, \&
  {Wisotzki}}]{Heymans+06}
{Heymans} C. {et~al.}, 2006, \mnras, 371, L60

\bibitem[{{Hilz} {et~al}\mbox{.}(2013){Hilz}, {Naab}, \& {Ostriker}}]{Hilz+13}
{Hilz} M., {Naab} T., {Ostriker} J.~P., 2013, \mnras, 429, 2924

\bibitem[{{Hopkins} {et~al}\mbox{.}(2008){Hopkins}, {Hernquist}, {Cox},
  {Dutta}, \& {Rothberg}}]{Hopkins+08_DELGN_I}
{Hopkins} P.~F., {Hernquist} L., {Cox} T.~J., {Dutta} S.~N.,
{Rothberg} B.,
  2008, \apj, 679, 156

\bibitem[{{Hopkins} {et~al}\mbox{.}(2009){Hopkins}, {Hernquist}, {Cox},
  {Keres}, \& {Wuyts}}]{Hopkins+09_DELGN_IV}
{Hopkins} P.~F., {Hernquist} L., {Cox} T.~J., {Keres} D., {Wuyts}
S., 2009,
  \apj, 691, 1424

\bibitem[{{Humphrey} \& {Buote}(2010)}]{Humphrey_Buote10}
{Humphrey} P.~J., {Buote} D.~A., 2010, \mnras, 403, 2143

\bibitem[{{Hyde} \& {Bernardi}(2009)}]{HB09_FP}
{Hyde} J.~B., {Bernardi} M., 2009, \mnras, 396, 1171

\bibitem[{{Khochfar} \& {Silk}(2006)}]{Khochfar_Silk06}
{Khochfar} S., {Silk} J., 2006, \apjl, 648, L21

\bibitem[{{Kochanek}(1991)}]{Kochanek91}
{Kochanek} C.~S., 1991, \apj, 373, 354

\bibitem[{{Komatsu} {et~al}\mbox{.}(2011){Komatsu}, {Smith}, {Dunkley},
  {Bennett}, {Gold}, {Hinshaw}, {Jarosik}, {Larson}, {Nolta}, {Page},
  {Spergel}, {Halpern}, {Hill}, {Kogut}, {Limon}, {Meyer}, {Odegard}, {Tucker},
  {Weiland}, {Wollack}, \& {Wright}}]{Komatsu+11_WMAP7}
{Komatsu} E. {et~al.}, 2011, \apjs, 192, 18

\bibitem[{{Koopmans} {et~al}\mbox{.}(2006){Koopmans}, {Treu}, {Bolton},
  {Burles}, \& {Moustakas}}]{Koopmans+06_SLACSIII}
{Koopmans} L.~V.~E., {Treu} T., {Bolton} A.~S., {Burles} S.,
{Moustakas} L.~A.,
  2006, \apj, 649, 599

\bibitem[{{La Barbera} \& {de Carvalho}(2009)}]{LaBarbera_deCarvalho09}
{La Barbera} F., {de Carvalho} R.~R., 2009, \apjl, 699, L76

\bibitem[{{La Barbera} {et~al}\mbox{.}(2010){La Barbera}, {de Carvalho}, {de La
  Rosa}, {Lopes}, {Kohl-Moreira}, \& {Capelato}}]{SPIDER-I}
{La Barbera} F., {de Carvalho} R.~R., {de La Rosa} I.~G., {Lopes}
P.~A.~A.,
  {Kohl-Moreira} J.~L., {Capelato} H.~V., 2010, \mnras, 408, 1313

\bibitem[{{La Barbera} {et~al}\mbox{.}(2013){La Barbera}, {Ferreras},
  {Vazdekis}, {de la Rosa}, {de Carvalho}, {Trevisan}, {Falc{\'o}n-Barroso}, \&
  {Ricciardelli}}]{LaBarbera+13_SPIDERVIII_IMF}
{La Barbera} F., {Ferreras} I., {Vazdekis} A., {de la Rosa} I.~G.,
{de
  Carvalho} R.~R., {Trevisan} M., {Falc{\'o}n-Barroso} J., {Ricciardelli} E.,
  2013, \mnras, 433, 3017

\bibitem[{{Lagattuta} {et~al}\mbox{.}(2010){Lagattuta}, {Fassnacht}, {Auger},
  {Marshall}, {Brada{\v c}}, {Treu}, {Gavazzi}, {Schrabback}, {Faure}, \&
  {Anguita}}]{Lagattuta+10}
{Lagattuta} D.~J. {et~al.}, 2010, \apj, 716, 1579

\bibitem[{{Macci{\`o}} {et~al}\mbox{.}(2008){Macci{\`o}}, {Dutton}, \& {van den
  Bosch}}]{Maccio+08}
{Macci{\`o}} A.~V., {Dutton} A.~A., {van den Bosch} F.~C., 2008,
\mnras, 391,
  1940

\bibitem[{{Mandelbaum} {et~al}\mbox{.}(2006){Mandelbaum}, {Seljak},
  {Kauffmann}, {Hirata}, \& {Brinkmann}}]{Mandelbaum+06}
{Mandelbaum} R., {Seljak} U., {Kauffmann} G., {Hirata} C.~M.,
{Brinkmann} J.,
  2006, \mnras, 368, 715

\bibitem[{{Marinoni} \& {Hudson}(2002)}]{MH02}
{Marinoni} C., {Hudson} M.~J., 2002, \apj, 569, 101

\bibitem[{{Moster} {et~al}\mbox{.}(2010){Moster}, {Somerville}, {Maulbetsch},
  {van den Bosch}, {Macci{\`o}}, {Naab}, \& {Oser}}]{Moster+10}
{Moster} B.~P., {Somerville} R.~S., {Maulbetsch} C., {van den
Bosch} F.~C.,
  {Macci{\`o}} A.~V., {Naab} T., {Oser} L., 2010, \apj, 710, 903

\bibitem[{{Napolitano} {et~al}\mbox{.}(2005){Napolitano}, {Capaccioli},
  {Romanowsky}, {Douglas}, {Merrifield}, {Kuijken}, {Arnaboldi}, {Gerhard}, \&
  {Freeman}}]{Napolitano+05}
{Napolitano} N.~R. {et~al.}, 2005, \mnras, 357, 691

\bibitem[{{Napolitano} {et~al}\mbox{.}(2010){Napolitano}, {Romanowsky}, \&
  {Tortora}}]{NRT10}
{Napolitano} N.~R., {Romanowsky} A.~J., {Tortora} C., 2010,
\mnras, 405, 2351

\bibitem[{{Navarro} {et~al}\mbox{.}(1996){Navarro}, {Frenk}, \&
  {White}}]{NFW96}
{Navarro} J.~F., {Frenk} C.~S., {White} S.~D.~M., 1996, \apj, 462,
563

\bibitem[{{Oguri} {et~al}\mbox{.}(2014){Oguri}, {Rusu}, \& {Falco}}]{Oguri+14}
{Oguri} M., {Rusu} C.~E., {Falco} E.~E., 2014, \mnras, 439, 2494

\bibitem[{{Padmanabhan} {et~al}\mbox{.}(2004){Padmanabhan}, {Seljak},
  {Strauss}, {Blanton}, {Kauffmann}, {Schlegel}, {Tremonti}, {Bahcall},
  {Bernardi}, {Brinkmann}, {Fukugita}, \& {Ivezi{\'c}}}]{Padmanabhan+04}
{Padmanabhan} N. {et~al.}, 2004, \na, 9, 329

\bibitem[{{Prugniel} \& {Simien}(1996)}]{PS96}
{Prugniel} P., {Simien} F., 1996, \aap, 309, 749

\bibitem[{{Prugniel} \& {Simien}(1997)}]{PS97}
{Prugniel} P., {Simien} F., 1997, \aap, 321, 111

\bibitem[{{Remus} {et~al}\mbox{.}(2013){Remus}, {Burkert}, {Dolag},
  {Johansson}, {Naab}, {Oser}, \& {Thomas}}]{Remus+13}
{Remus} R.-S., {Burkert} A., {Dolag} K., {Johansson} P.~H., {Naab}
T., {Oser}
  L., {Thomas} J., 2013, \apj, 766, 71

\bibitem[{{Roche} {et~al}\mbox{.}(2010){Roche}, {Bernardi}, \&
  {Hyde}}]{Roche+10}
{Roche} N., {Bernardi} M., {Hyde} J., 2010, \mnras, 407, 1231

\bibitem[{{Rudnick} {et~al}\mbox{.}(2009){Rudnick}, {von der Linden},
  {Pell{\'o}}, {Arag{\'o}n-Salamanca}, {Marchesini}, {Clowe}, {De Lucia},
  {Halliday}, {Jablonka}, {Milvang-Jensen}, {Poggianti}, {Saglia}, {Simard},
  {White}, \& {Zaritsky}}]{Rudnick+09}
{Rudnick} G. {et~al.}, 2009, \apj, 700, 1559

\bibitem[{{Ruff} {et~al}\mbox{.}(2011){Ruff}, {Gavazzi}, {Marshall}, {Treu},
  {Auger}, \& {Brault}}]{Ruff+11}
{Ruff} A.~J., {Gavazzi} R., {Marshall} P.~J., {Treu} T., {Auger}
M.~W.,
  {Brault} F., 2011, \apj, 727, 96

\bibitem[{{Rusin} {et~al}\mbox{.}(2003){Rusin}, {Kochanek}, \&
  {Keeton}}]{Rusin+03}
{Rusin} D., {Kochanek} C.~S., {Keeton} C.~R., 2003, \apj, 595, 29

\bibitem[{{Ruszkowski} \& {Springel}(2009)}]{RS09}
{Ruszkowski} M., {Springel} V., 2009, \apj, 696, 1094

\bibitem[{{Saglia} {et~al}\mbox{.}(2010){Saglia}, {S{\'a}nchez-Bl{\'a}zquez},
  {Bender}, {Simard}, {Desai}, {Arag{\'o}n-Salamanca}, {Milvang-Jensen},
  {Halliday}, {Jablonka}, {Noll}, {Poggianti}, {Clowe}, {De Lucia},
  {Pell{\'o}}, {Rudnick}, {Valentinuzzi}, {White}, \& {Zaritsky}}]{Saglia+10}
{Saglia} R.~P. {et~al.}, 2010, \aap, 524, A6

\bibitem[{{Salpeter}(1955)}]{Salpeter55}
{Salpeter} E.~E., 1955, \apj, 121, 161

\bibitem[{{Schneider} {et~al}\mbox{.}(2012){Schneider}, {Smith}, {Macci{\`o}},
  \& {Moore}}]{Schneider+12}
{Schneider} A., {Smith} R.~E., {Macci{\`o}} A.~V., {Moore} B.,
2012, \mnras,
  424, 684

\bibitem[{{Shankar} \& {Bernardi}(2009)}]{Shankar_Bernardi09}
{Shankar} F., {Bernardi} M., 2009, \mnras, 396, L76

\bibitem[{{Shankar} {et~al}\mbox{.}(2010){Shankar}, {Marulli}, {Bernardi},
  {Dai}, {Hyde}, \& {Sheth}}]{Shankar+10}
{Shankar} F., {Marulli} F., {Bernardi} M., {Dai} X., {Hyde} J.~B.,
{Sheth}
  R.~K., 2010, \mnras, 403, 117

\bibitem[{{Shetty} \& {Cappellari}(2014)}]{Shetty_Cappellari14}
{Shetty} S., {Cappellari} M., 2014, \apjl, 786, L10

\bibitem[{{Sonnenfeld} {et~al}\mbox{.}(2013){Sonnenfeld}, {Treu}, {Gavazzi},
  {Suyu}, {Marshall}, {Auger}, \& {Nipoti}}]{Sonnenfeld+13_SL2S_IV}
{Sonnenfeld} A., {Treu} T., {Gavazzi} R., {Suyu} S.~H., {Marshall}
P.~J.,
  {Auger} M.~W., {Nipoti} C., 2013, \apj, 777, 98

\bibitem[{{Sparks} \& {Jorgensen}(1993)}]{Sparks_Jorgensen93}
{Sparks} W.~B., {Jorgensen} I., 1993, \aj, 105, 1753

\bibitem[{{Spiniello} {et~al}\mbox{.}(2011){Spiniello}, {Koopmans}, {Trager},
  {Czoske}, \& {Treu}}]{Spiniello+11}
{Spiniello} C., {Koopmans} L.~V.~E., {Trager} S.~C., {Czoske} O.,
{Treu} T.,
  2011, \mnras, 417, 3000

\bibitem[{{Spiniello} {et~al}\mbox{.}(2012){Spiniello}, {Trager}, {Koopmans},
  \& {Chen}}]{Spiniello+12}
{Spiniello} C., {Trager} S.~C., {Koopmans} L.~V.~E., {Chen} Y.~P.,
2012, \apjl,
  753, L32

\bibitem[{{Swindle} {et~al}\mbox{.}(2011){Swindle}, {Gal}, {La Barbera}, \& {de
  Carvalho}}]{SPIDER-V}
{Swindle} R., {Gal} R.~R., {La Barbera} F., {de Carvalho} R.~R.,
2011, \aj,
  142, 118

\bibitem[{{Thomas} {et~al}\mbox{.}(2013){Thomas}, {Steele}, {Maraston},
  {Johansson}, {Beifiori}, {Pforr}, {Str{\"o}mb{\"a}ck}, {Tremonti}, {Wake},
  {Bizyaev}, {Bolton}, {Brewington}, {Brownstein}, {Comparat}, {Kneib},
  {Malanushenko}, {Malanushenko}, {Oravetz}, {Pan}, {Parejko}, {Schneider},
  {Shelden}, {Simmons}, {Snedden}, {Tanaka}, {Weaver}, \&
  {Yan}}]{Thomas+13_BOSS}
{Thomas} D. {et~al.}, 2013, \mnras, 431, 1383

\bibitem[{{Thomas} {et~al}\mbox{.}(2007){Thomas}, {Saglia}, {Bender}, {Thomas},
  {Gebhardt}, {Magorrian}, {Corsini}, \& {Wegner}}]{ThomasJ+07}
{Thomas} J., {Saglia} R.~P., {Bender} R., {Thomas} D., {Gebhardt}
K.,
  {Magorrian} J., {Corsini} E.~M., {Wegner} G., 2007, \mnras, 382, 657

\bibitem[{{Thomas} {et~al}\mbox{.}(2009){Thomas}, {Saglia}, {Bender}, {Thomas},
  {Gebhardt}, {Magorrian}, {Corsini}, \& {Wegner}}]{ThomasJ+09}
{Thomas} J., {Saglia} R.~P., {Bender} R., {Thomas} D., {Gebhardt}
K.,
  {Magorrian} J., {Corsini} E.~M., {Wegner} G., 2009, \apj, 691, 770

\bibitem[{{Thomas} {et~al}\mbox{.}(2011){Thomas}, {Saglia}, {Bender}, {Thomas},
  {Gebhardt}, {Magorrian}, {Corsini}, {Wegner}, \& {Seitz}}]{ThomasJ+11}
{Thomas} J. {et~al.}, 2011, \mnras, 415, 545

\bibitem[{{Tortora} {et~al}\mbox{.}(2012){Tortora}, {La Barbera}, {Napolitano},
  {de Carvalho}, \& {Romanowsky}}]{SPIDER-VI}
{Tortora} C., {La Barbera} F., {Napolitano} N.~R., {de Carvalho}
R.~R.,
  {Romanowsky} A.~J., 2012, \mnras, 425, 577

\bibitem[{{Tortora} {et~al}\mbox{.}(2010{\natexlab{a}}){Tortora}, {Napolitano},
  {Cardone}, {Capaccioli}, {Jetzer}, \& {Molinaro}}]{Tortora+10CG}
{Tortora} C., {Napolitano} N.~R., {Cardone} V.~F., {Capaccioli}
M., {Jetzer}
  P., {Molinaro} R., 2010{\natexlab{a}}, \mnras, 407, 144

\bibitem[{{Tortora} {et~al}\mbox{.}(2009){Tortora}, {Napolitano}, {Romanowsky},
  {Capaccioli}, \& {Covone}}]{Tortora+09}
{Tortora} C., {Napolitano} N.~R., {Romanowsky} A.~J., {Capaccioli}
M., {Covone}
  G., 2009, \mnras, 396, 1132

\bibitem[{{Tortora} {et~al}\mbox{.}(2010{\natexlab{b}}){Tortora}, {Napolitano},
  {Romanowsky}, \& {Jetzer}}]{Tortora+10lensing}
{Tortora} C., {Napolitano} N.~R., {Romanowsky} A.~J., {Jetzer} P.,
  2010{\natexlab{b}}, \apjl, 721, L1

\bibitem[{{Tortora} {et~al}\mbox{.}(2014){Tortora}, {Romanowsky}, {Cardone},
  {Napolitano}, \& {Jetzer}}]{Tortora+14_MOND}
{Tortora} C., {Romanowsky} A.~J., {Cardone} V.~F., {Napolitano}
N.~R., {Jetzer}
  P., 2014, \mnras, 438, L46

\bibitem[{{Tortora} {et~al}\mbox{.}(2013){Tortora}, {Romanowsky}, \&
  {Napolitano}}]{TRN13_SPIDER_IMF}
{Tortora} C., {Romanowsky} A.~J., {Napolitano} N.~R., 2013, \apj,
765, 8

\bibitem[{{Treu} {et~al}\mbox{.}(2010){Treu}, {Auger}, {Koopmans}, {Gavazzi},
  {Marshall}, \& {Bolton}}]{Treu+10}
{Treu} T., {Auger} M.~W., {Koopmans} L.~V.~E., {Gavazzi} R.,
{Marshall} P.~J.,
  {Bolton} A.~S., 2010, \apj, 709, 1195

\bibitem[{{Treu} \& {Koopmans}(2004)}]{TK04}
{Treu} T., {Koopmans} L.~V.~E., 2004, \apj, 611, 739

\bibitem[{{Trujillo} {et~al}\mbox{.}(2004){Trujillo}, {Burkert}, \&
  {Bell}}]{TBB04}
{Trujillo} I., {Burkert} A., {Bell} E.~F., 2004, \apjl, 600, L39

\bibitem[{{Trujillo} {et~al}\mbox{.}(2011){Trujillo}, {Ferreras}, \& {de La
  Rosa}}]{Trujillo+11}
{Trujillo} I., {Ferreras} I., {de La Rosa} I.~G., 2011, \mnras,
415, 3903

\bibitem[{{Trujillo} {et~al}\mbox{.}(2006){Trujillo}, {F{\"o}rster Schreiber},
  {Rudnick}, {Barden}, {Franx}, {Rix}, {Caldwell}, {McIntosh}, {Toft},
  {H{\"a}ussler}, {Zirm}, {van Dokkum}, {Labb{\'e}}, \& {}}]{Trujillo+06}
{Trujillo} I. {et~al.}, 2006, \apj, 650, 18

\bibitem[{{Valentinuzzi} {et~al}\mbox{.}(2010{\natexlab{a}}){Valentinuzzi},
  {Fritz}, {Poggianti}, {Cava}, {Bettoni}, {Fasano}, {D'Onofrio}, {Couch},
  {Dressler}, {Moles}, {Moretti}, {Omizzolo}, {Kj{\ae}rgaard}, {Vanzella}, \&
  {Varela}}]{Valentinuzzi+10_WINGS}
{Valentinuzzi} T. {et~al.}, 2010{\natexlab{a}}, \apj, 712, 226

\bibitem[{{Valentinuzzi} {et~al}\mbox{.}(2010{\natexlab{b}}){Valentinuzzi},
  {Poggianti}, {Saglia}, {Arag{\'o}n-Salamanca}, {Simard},
  {S{\'a}nchez-Bl{\'a}zquez}, {D'onofrio}, {Cava}, {Couch}, {Fritz}, {Moretti},
  \& {Vulcani}}]{Valentinuzzi+10_EDisCS}
{Valentinuzzi} T. {et~al.}, 2010{\natexlab{b}}, \apjl, 721, L19

\bibitem[{{van den Bosch} {et~al}\mbox{.}(2007){van den Bosch}, {Yang}, {Mo},
  {Weinmann}, {Macci{\`o}}, {More}, {Cacciato}, {Skibba}, \& {Kang}}]{vdB+07}
{van den Bosch} F.~C. {et~al.}, 2007, \mnras, 376, 841

\bibitem[{{van Dokkum} \& {Franx}(2001)}]{vanDokkum_Franx01}
{van Dokkum} P.~G., {Franx} M., 2001, \apj, 553, 90

\bibitem[{{van Dokkum} {et~al}\mbox{.}(2010){van Dokkum}, {Whitaker},
  {Brammer}, {Franx}, {Kriek}, {Labb{\'e}}, {Marchesini}, {Quadri}, {Bezanson},
  {Illingworth}, {Muzzin}, {Rudnick}, {Tal}, \& {Wake}}]{vanDokkum+10}
{van Dokkum} P.~G. {et~al.}, 2010, \apj, 709, 1018

\bibitem[{{Vazdekis} {et~al}\mbox{.}(2012){Vazdekis}, {Ricciardelli},
  {Cenarro}, {Rivero-Gonz{\'a}lez}, {D{\'{\i}}az-Garc{\'{\i}}a}, \&
  {Falc{\'o}n-Barroso}}]{Vazdekis+12}
{Vazdekis} A., {Ricciardelli} E., {Cenarro} A.~J.,
{Rivero-Gonz{\'a}lez} J.~G.,
  {D{\'{\i}}az-Garc{\'{\i}}a} L.~A., {Falc{\'o}n-Barroso} J., 2012, \mnras,
  3156

\bibitem[{{Vulcani} {et~al}\mbox{.}(2014){Vulcani}, {Bamford},
  {H{\"a}u{\ss}ler}, {Vika}, {Rojas}, {Agius}, {Baldry}, {Bauer}, {Brown},
  {Driver}, {Graham}, {Kelvin}, {Liske}, {Loveday}, {Popescu}, {Robotham}, \&
  {Tuffs}}]{Vulcani+14}
{Vulcani} B. {et~al.}, 2014, \mnras, 441, 1340

\bibitem[{{Vulcani} {et~al}\mbox{.}(2011){Vulcani}, {Poggianti},
  {Arag{\'o}n-Salamanca}, {Fasano}, {Rudnick}, {Valentinuzzi}, {Dressler},
  {Bettoni}, {Cava}, {D'Onofrio}, {Fritz}, {Moretti}, {Omizzolo}, \&
  {Varela}}]{Vulcani+11}
{Vulcani} B. {et~al.}, 2011, \mnras, 412, 246

\bibitem[{{Wegner} {et~al}\mbox{.}(2012){Wegner}, {Corsini}, {Thomas},
  {Saglia}, {Bender}, \& {Pu}}]{Wegner+12}
{Wegner} G.~A., {Corsini} E.~M., {Thomas} J., {Saglia} R.~P.,
{Bender} R., {Pu}
  S.~B., 2012, \aj, 144, 78

\bibitem[{{Weidner} {et~al}\mbox{.}(2013){Weidner}, {Ferreras}, {Vazdekis}, \&
  {La Barbera}}]{Weidner+13_giant_ell}
{Weidner} C., {Ferreras} I., {Vazdekis} A., {La Barbera} F., 2013,
\mnras, 435,
  2274

\bibitem[{{White} {et~al}\mbox{.}(2005){White}, {Clowe}, {Simard}, {Rudnick},
  {De Lucia}, {Arag{\'o}n-Salamanca}, {Bender}, {Best}, {Bremer}, {Charlot},
  {Dalcanton}, {Dantel}, {Desai}, {Fort}, {Halliday}, {Jablonka}, {Kauffmann},
  {Mellier}, {Milvang-Jensen}, {Pell{\'o}}, {Poggianti}, {Poirier},
  {Rottgering}, {Saglia}, {Schneider}, \& {Zaritsky}}]{White+05}
{White} S.~D.~M. {et~al.}, 2005, \aap, 444, 365

\bibitem[{{Wu} {et~al}\mbox{.}(2014){Wu}, {Gerhard}, {Naab}, {Oser},
  {Martinez-Valpuesta}, {Hilz}, {Churazov}, \& {Lyskova}}]{Wu+14}
{Wu} X., {Gerhard} O., {Naab} T., {Oser} L., {Martinez-Valpuesta}
I., {Hilz}
  M., {Churazov} E., {Lyskova} N., 2014, \mnras, 438, 2701

\end{thebibliography}

\end{document}